\documentstyle[12pt,epsf]{article}
\newcommand{\beq}{\begin{equation}}
\newcommand{\eeq}{\end{equation}}
\newcommand{\beqa}{\begin{eqnarray}}
\newcommand{\eeqa}{\end{eqnarray}}

\newcommand{\barr}[1]{\not\mathrel #1}
\newcommand{\Tr}{{\rm Tr}}
\newcommand{\dfrac}{\displaystyle \frac}

\begin{document}
\newpage
\baselineskip 16pt plus 2pt minus 2pt

\thispagestyle{empty}

\par
\topmargin=-1cm      

{ \small

\hfill{KFA-IKP(TH)-1997-05} 

\hfill{LPT 97-03}

}

\vspace{30.0pt}

\begin{centering}
{\Large\bf The reaction 
$\pi N \to \pi \pi N$ above threshold in \\[0.3em]
chiral perturbation theory}\\

\vspace{40.0pt}
{{\bf V.~Bernard}$^1$,
{\bf N.~Kaiser}$^2$,
{\bf Ulf-G.~Mei{\ss}ner}$^3$}\\
\vspace{20.0pt}
{\sl $^{1}$Laboratoire de Physique Th\'eorique} \\
{\sl Universit\'e Louis Pasteur, F-67037 Strasbourg Cedex 2, France} \\
{\it E-mail address: bernard@crnhp4.in2p3.fr}\\
\vspace{15.0pt}
{\sl $^{2}$Physik Department T39, TU M\"unchen, D-85747 Garching, Germany}\\
{\it E-mail address: nkaiser@physik.tu-muenchen.de}\\ 
\vspace{15.0pt}
{\sl $^{3}$Institut f\"ur Kernphysik, Forschungszentrum J\"ulich, 
D-52425 J\"ulich, Germany} \\
{\it E-mail address: Ulf-G.Meissner@kfa-juelich.de}\\
\end{centering}
\vspace{40.0pt}
\begin{center}
\begin{abstract}
\noindent Single pion production off nucleons is studied in
the framework of relativistic baryon chiral perturbation theory 
at tree level with the inclusion of the terms from the dimension two
effective pion--nucleon Lagrangian. The five appearing low--energy
constants are fixed from pion--nucleon scattering data. Despite
the simplicity of the approach, most  of the existing
data for total and differential cross sections as well as for the
angular correlation functions for incoming pion kinetic energies up to 400~MeV
can be satisfactorily described.
\end{abstract}

\vspace*{70pt}
\small{PACS nos.: 25.80.Hp , 12.39.Fe , 11.30.Rd}

\vfill
\end{center}

\newpage

\section{Introduction}

Single pion production off nucleons has been at the center of numerous
experimental and theoretical investigations since many years. One of
the original motivations of these works was the observation that
the elusive pion--pion threshold S--wave interaction can be deduced
from the pion--pole graph contribution. A whole series of precision
experiments at PSI, TRIUMF and CERN has been performed over the last
decade and there is still on--going activity. On the theoretical side,
chiral perturbation theory has emerged as a precision tool in low
energy hadron physics. Beringer considered the reaction $\pi N \to \pi
\pi N$ to lowest order in chiral perturbation theory
\cite{bering}. Low--energy theorems for the threshold amplitudes $D_1$
and $D_2$\footnote{These are related to the more commonly used 
${\cal A}_{10}$ and ${\cal A}_{32}$ by ${\cal A}_{32} = \sqrt{10}D_1$
and ${\cal A}_{10} = -2D_1 - 3D_2$.} were derived in \cite{bkmplb}.
These are free of unknown parameters and not sensitive to the $\pi
\pi$ interaction beyond tree level. 
A direct comparison with the threshold data for
the channel $\pi^+ p \to \pi^+ \pi^+ n$, which is only sensitive to $D_1$,
lead to a very satisfactory description whereas in case of the process
$\pi^- p \to \pi^0 \pi^0 n$, which is only sensitive to $D_2$, sizeable 
deviations were found for the total cross sections near threshold. These were 
originally attributed to the strong pionic final--state interactions in the
$I_{\pi\pi}=0$ channel. However, this conjecture turned out to be incorrect 
when a complete higher order calculation of the threshold amplitudes $D_{1,2}$
was performed \cite{bkmppn}. In that paper, we investigated the relation
between the threshold amplitudes $D_1$ and $D_2$ for the reaction $\pi N \to
\pi \pi N$ and the $\pi\pi$ S--wave scattering lengths $a_0^0$ and $a_0^2$
in the framework of heavy baryon chiral perturbation theory to second
order in small momenta, i.e. the pion mass (which is the only small
parameter at threshold). The pion loop and pionic counterterm 
corrections  only start  contributing to the $\pi\pi N$ threshold amplitudes at
second order. One of these counterterms, proportional to the low--energy
constant $\ell_3$, eventually allows to
measure the scalar quark condensate, i.e. the strength of the spontaneous 
chiral symmetry breaking in QCD. However, at that order, the largest 
contribution stems indeed from insertions of the dimension two chiral
pion--nucleon Lagrangian, which is characterized by a few coupling constants
called $c_i$.  In particular this is the case for the amplitude $D_2$. To be 
specific, consider the threshold  amplitudes $D_{1,2}$ calculated
from the relativistic Born graphs (with lowest order vertices) and the
relativistic $c_i$--terms expanded  to second order in the pion mass. This 
gives 
\beqa \label{D12}
D_1^{{\rm Born}} + D_1^{c_i} &=& (2.33 + 0.24 \pm 0.10) \, {\rm
    fm}^3 = (2.57 \pm 0.10) \, {\rm  fm}^3 \,\,, \\
D_2^{{\rm Born}} + D_2^{c_i} &=& (-6.61 - 2.85 \pm 0.06) \, {\rm
    fm}^3 = (-9.46 \pm 0.06) \, {\rm  fm}^3 \,\, ,
\eeqa
which are within 15\% and 5\% off the empirical values, $D_1^{\exp} =(2.26 \pm
0.10)\,$fm$^3$ and $D_2^{\exp} = (-9.05 \pm 0.36)\,$fm$^3$,
respectively.  It appears therefore natural to extend the same
calculation above threshold and to compare to the large body of data
for the various reaction channels that exist. It was already shown by
Beringer \cite{bering} that taking simply the relativistic Born terms
does indeed not suffice to describe the total cross section data for
incoming pion energies up to 400~MeV in most channels. Such a failure 
can be expected from the threshold expansion of $D_2$, where the Born terms 
only amount to 75\% of the empirical value. We therefore expect that the
inclusion of the dimension two operators, which clearly improves the
prediction for $D_2$ at threshold, will lead to a better description of
the above threshold data. In particular, it will tell to which extent
loop effects are necessary (and thus testing the sensitivity to the
pion--pion interaction beyond tree level) and to which extent one has to
incorporate resonance degrees of freedom like the Roper and the
$\Delta$--isobar as well as heavier mesons ($\sigma, \rho,\omega $) 
as dynamical degrees of freedom (as it is done in
many models, see e.g. \cite{oset} \cite{jaeck}). Our calculation is in
spirit closest to the one of ref.\cite{jemi}, in which Beringer's Born
terms where supplemented by explicit $\Delta$ and Roper (tree) contributions.
Clearly, the inclusion of the resonances as done in that paper is not based on 
a consistent power counting scheme but rather it is argued that
phenomenology demands the extension of the effective Lagrangian to include
these higher mass states. It will be of particular importance to directly 
compare our results to the ones presented there for the abovementioned reasons.
Since we will only calculate tree level diagrams with at most one insertion
from the dimension two pion--nucleon Lagrangian, it is advantageous to treat 
the nucleons as fully relativistic Dirac fields, since then one automatically
generates all $1/m$ suppressed terms with fixed coefficients. Here,
 $m$ denotes the 
nucleon mass. Only at next order, when loop graphs have to be included, the 
problems due to the additional large mass scale given by the nucleon mass
appear \cite{gss}. Therefore, the calculation presented here can not 
easily be extended to higher orders in the chiral expansion. At this point,
we stress that the amplitudes calculated from the tree graphs as done here
are, of course, purely real. How severe this approximation is can only be
judged when a full one loop calculation above threshold has been performed.

This paper is organized as follows. Section 2 contains some formal
aspects including the definition of the T--matrix and of the
pertinent observables. In section 3 we briefly review the
effective Lagrangian underlying the calculation and
discuss the tree level diagrams to be evaluated.
Section 4 contains the results and discussions thereof.  
The appendices include more details about the kinematics, some
lengthy formula appearing in the expressions for the observables and
the explicit expressions for the invariant amplitudes.

\section{Single pion production: Formal aspects}

In this section, we outline the basic framework concerning the
reaction $\pi N \to \pi \pi N$ above threshold. Some of the
lengthy formulae are relegated to the appendices.

\subsection{T-matrix and invariant amplitudes}

We seek the T--matrix element for the process
\beq
\pi^a(k)+N(p_1)\to\pi^b(q_1)+\pi^c(q_2)+N(p_2) \,\,\, ,
\eeq
with $N$ denoting a nucleon (neutron or proton) and $\pi^a$ a pion of
(cartesian) isospin $a$. This process is characterized by  five independent 
Mandelstam variables,
\begin{equation} \label{Mand} 
s=(p_1+k)^2,\quad s_1=(q_1+p_2)^2,\quad s_2=(q_2+p_2)^2,\quad
t_1=(q_1-k)^2,\quad t_2=(q_2-k)^2 \, .
\end{equation}
All scalar products between the various momenta can be expressed 
in terms of these, see appendix~A, and at threshold 
$(\vec{q}_1 = \vec{q}_2 = 0$, in the center--of--mass frame), we have
\begin{equation} 
s^{\rm thr}=(m+2M_\pi)^2,\quad s_1^{\rm thr}=s_2^{\rm
thr}=(m+M_\pi)^2,\quad t_1^{\rm thr}= t_2^{\rm thr}=-M_\pi^2\,
{2m+M_\pi \over m +2M_\pi} \,\, . 
\end{equation}
The $\pi N\to \pi \pi N$ transition matrix element can be expressed
in terms of four invariant amplitudes, denoted
$F_i(s,s_1,s_2,t_1,t_2)\, (i=1,2,3,4)$, 
\begin{equation} 
T = i \, \overline u_2 \gamma_5\bigl[ F_1 + (\barr q_1+
\barr q_2) F_2+(\barr q_1-\barr q_2) F_3+ (\barr q_1 \barr q_2-\barr q_2 
\barr q_1)  F_4 \bigr] u_1 \,\, .
\end{equation}
with $u_{1,2}$ the Dirac spinor for the in- and outgoing nucleon, 
respectively. 
In the complete relativistic tree calculation to order $q^2$, these are 
rational real functions of the five Mandelstam variables. 
The isospin decomposition of the invariant amplitudes reads (pulling
out a common prefactor composed of coupling constants)
\begin{equation} 
F_i = {g_A \over 8F_\pi^3} \bigg[ \tau^a \delta^{bc}\, B_i^1 +
(\tau^b \delta^{ac} +\tau^c \delta^{ab})\, B_i^2 +(\tau^b \delta^{ac}-\tau^c 
\delta^{ab})\, B_i^3 +i\epsilon^{abc}\, B_i^4 \bigg] \,\, , 
\end{equation}
with $g_A$ the nucleon axial--vector coupling  constant and $F_\pi$
the pion decay constant, respectively. The crossing properties of the
isospin amplitudes can be readily deduced.  Crossing the two out-going
pion  lines, $(q_1,b) \leftrightarrow (q_2,c)$, amounts to 
\begin{equation} 
B_i^j(s,s_1,s_2,t_1,t_2) = \epsilon_i \epsilon_j B_i^j(s,s_2,
s_1,t_2,t_1)\,, \quad \epsilon_{1,2} = +1\,, \quad \epsilon_{3,4} = -1
\,\, .
\end{equation}    
In the physical reaction channels the relevant amplitudes are given by 
\begin{equation} 
\tilde F_i = {g_A \over 8F_\pi^3} \sum_{j=1}^4 \eta_j B_i^j \,\,,
\end{equation}
with $\eta_j$ channel dependent isospin factors. We calculate the amplitudes
$B^j_i$ in the isospin limit with the charged pion
mass $(M_\pi = 139.57$~MeV) and the proton mass, $m = 938.27$~MeV, in order. 
Isospin breaking is done by shifting the kinetic energy of the
incoming pion, $T_\pi$, from the isospin symmetric threshold 
\beq
T_\pi^{{\rm thr,iso}}= M_\pi \bigg(1+{3M_\pi \over 2m}\bigg) =170.71\,\, {\rm
  MeV}
\eeq 
to the correct threshold. The pertinent isospin coefficients $\eta_j$ and 
numerical values for this shift, called $\delta T_\pi$, for the five
experimentally accessible reaction channels are
\beqa
\pi^+p\to \pi^+ \pi^+ n &:& \quad \eta_1=0\,, \eta_2=2\sqrt{2}\,,
\eta_3=0\,,  \eta_4=0\,, \delta T_\pi = +1.68\, {\rm MeV} \\ 
\pi^+p\to \pi^+ \pi^0 p &:& \quad \eta_1=0\,, \eta_2=1 \,,
\eta_3=-1\,,  \eta_4=1\,, \delta T_\pi =  -5.95\, {\rm MeV} \\ 
\pi^-p\to\pi^+ \pi^- n &:& \quad\eta_1=\sqrt{2}\,,\eta_2=\sqrt{2}
\,, \eta_3=\sqrt{2}\,, \eta_4=0\,,\delta T_\pi = +1.68\, {\rm MeV} \\ 
\pi^-p\to \pi^0 \pi^0 n &:& \quad \eta_1=\sqrt{2} \,, \eta_2=0\,,
\eta_3=0\,,  \eta_4=0\,,  \delta T_\pi = -10.21\, {\rm MeV} \\
\pi^-p\to \pi^0 \pi^- p &:& \quad \eta_1=0\,, \eta_2=1 \,,
\eta_3=1\,,  \eta_4=1\,, \delta T_\pi = -5.95\, {\rm MeV} 
\, \, . 
\eeqa

\subsection{Observables}

The total cross section is a four--dimensional integral over a quadratic form
in  the amplitudes $\tilde F_i$, 
\begin{equation} \label{xstot}
\sigma_{\rm tot}(T_\pi)= { {\cal S} \over (4\pi)^4 m
\sqrt{T_\pi(T_\pi+2M_\pi)}} \int\int_{ z^2<1} d\omega_1d\omega_2 \int_{-1}^1
dx_1 \int_0^\pi d \phi \sum_{i,j=1}^4 y_{ij}\tilde F^*_i\tilde F_j \,\, .
\end{equation} 
The weight functions $y_{ij}$ are given in appendix~A, and ${\cal S}$
is a Bose symmetry factor, ${\cal S}=1/2$ for identical pions in the
final state and ${\cal S}=1$ otherwise.
The center--of--mass (CMS) kinematics is
\begin{equation} 
s = (m+M_\pi)^2 + 2m T_\pi\,,\quad k_0 ={s-m^2+M_\pi^2 \over 2 \sqrt{s}}\,,
\quad  |\vec k|  = \sqrt{k_0^2-M_\pi^2}\,, \quad |\vec q_i|  =\sqrt{\omega_i^2
-M_\pi^2} \,\, , \end{equation}
\begin{equation} s_i  = s-2\sqrt{s}\,\omega_{3-i}+M_\pi^2\,, \quad t_i  = 2(M_
\pi^2 - k_0 \,\omega_i + |\vec k||\vec q_i| x_i) \quad (i=1,2) \,\, , 
\end{equation}
\begin{equation} x_2 = x_1z+\sqrt{(1-x_1^2)(1-z^2)}\cos \phi
\,\, ,
\end{equation}
\begin{equation} z \, |\vec q_1||\vec q_2| = \omega_1 \omega_2 - \sqrt{s} (
\omega_1+\omega_2) + M_\pi^2 +{1\over2}(s-m^2) \,\, .
\end{equation}
Here, $\phi$ is the (auxiliary) angle between the planes spanned by $\vec q_1$
and $\vec k$ as well as $\vec q_1$ and $\vec q_2$. In accordance with the
experimentalists convention, we have chosen the coordinate frame such that the
incoming pion momentum $\vec k$ defines the z--direction, whereas $\vec q_1$,
the momentum of $\pi^b$ lies in the  xz--plane. The polar angles $\theta_1$
and $\theta_2$ of the outgoing pions (with $x_i = \cos\theta_i$) are in general
non--vanishing and so is the azimuthal angle $\phi_2$ of $\pi^c$. By
construction, the azimuthal angle of $\pi^b$ is zero.

\noindent The double differential cross section in the CMS depends on $T_\pi$,
$\omega_1$, and $\theta_1$. For this more exclusive quantity the energy 
$\omega_1$ and the polar angle $\theta_1$ of  $\pi^b$ are detected,  
\begin{equation} \label{xsdd}
{d^2 \sigma \over d \omega_1 d\Omega_1 } = {2 {\cal S} \over
(4\pi)^5 \sqrt{s} |\vec k| } \int_{\omega_2^-}^{\omega_2^+} d\omega_2
\int_0^\pi d\phi  \sum_{i,j=1}^4 y_{ij}\tilde F^*_i\tilde F_j 
\end{equation}
with  the boundaries of the three--body phase space given by
\begin{eqnarray} 
\omega_2^\pm &=& {1 \over 2(s-2\sqrt{s} \omega_1 +M_\pi^2)}
\Big[ (\sqrt{s}-\omega_1)(s-2\sqrt{s} \omega_1 -m^2+2M_\pi^2) \nonumber \\ & &
\pm |\vec q_1\,| \sqrt{(s-2\sqrt{s} \omega_1 -m^2)^2 -4m^2M^2_\pi} \Big] \, . 
\end{eqnarray}

\noindent Also measured are triple differential cross sections. These
depend on $T_\pi$, $\omega_1$, $\theta_1$,
$\theta_2$, and $ \phi_2$. The two angles of $\pi^c$ are measured in addition,
thus the three--particle final state is kinematically completely determined,
\begin{equation} \label{xstd}
{d^3 \sigma \over d\omega_1 d\Omega_1  d\Omega_2}
= { |\vec q_1||\vec q_2| {\cal S} \over (4\pi)^5 \sqrt{s}|\vec k| \tilde{E}_2}
\,\sum_{i,j=1}^4 y_{ij}\tilde F^*_i\tilde F_j  \,\, ,
\end{equation}
with $ s,\, s_1,\,s_2,\,t_1,\,t_2$ as before and 
\begin{equation} \tilde E_2 = E_2\biggl(1+{\partial E_2\over \partial
\omega_2}\biggr) = {\omega_2({1\over2}(s-m^2)-\sqrt{s}\,\omega_1)+M_\pi^2(
\omega_1 +\omega_2 -\sqrt{s})\over \omega_2^2-M_\pi^2} \,\, ,
\end{equation} 
with $E_2$ the final state nucleon energy. The cosine of the angle between
$\vec q_1$ and $\vec q_2$ is given by  
\begin{equation} z = \cos \theta_1\cos \theta_2+ \sin \theta_1\sin \theta_2
\cos \phi_2   \,\, ,   \end{equation}
and it is needed to express the energy of $\pi^c$ as 
\begin{eqnarray} \omega_2 &=& {1\over 2[(\sqrt{s}-\omega_1)^2-z^2|\vec q_1|^2]}
\biggl\{ (\sqrt{s}-\omega_1)(s-2\sqrt{s}\,\omega_1-m^2+2M_\pi^2) \nonumber \\ &
& - z |\vec q_1| \sqrt{(s-2\sqrt{s}\,\omega_1-m^2)^2-4
  M_\pi^2(m^2+(1-z^2)|\vec q_1|^2)}\biggr\} \,\, .
\end{eqnarray}
{}From the triple and double differential cross sections, one defines the 
angular correlation function $W$ at fixed beam energy $T_\pi$,
\begin{equation}W(\omega_1,\theta_1,\theta_2,\phi_2) = 4\pi {d^3 \sigma \over
d\omega_1 d\Omega_1 d\Omega_2 } \Bigg/ {d^2\sigma \over d\omega_1 d\Omega_1} 
\,\, .  \end{equation}
This completes the necessary formalism for our calculation.


\section{Calculation of the tree level amplitudes}

In this section, we briefly discuss the effective chiral Lagrangian
underlying our calculations. For
more details we refer to the review \cite{bkmrev} and to the paper 
\cite{bkmppn}. We then show and discuss the various classes of tree
diagrams contributing to the process $\pi N \to \pi \pi N$ to second
order in the chiral expansion.

\subsection{Effective Lagrangian}

At low energies, the relevant degrees of freedom are hadrons, in 
particular the Goldstone bosons linked to the spontaneous chiral symmetry
breaking. We consider here the two flavor case and thus deal with the
triplet of pions, collected in the matrix $U(x) = u^2(x)$.
It is straightforward to build an effective Lagrangian
to describe their interactions, called ${\cal L}_{\pi\pi}$. This Lagrangian
admits a dual expansion in small (external) momenta and quark (meson) 
masses as detailed below. Matter fields such as nucleons can also be 
included in the effective field theory based on the
familiar notions of non--linearly realized chiral symmetry.
The pertinent effective Lagrangian is called
${\cal L}_{\pi N}$,
consisting of terms with  exactly one nucleon in the initial 
and the final state. The various terms contributing to a process
under consideration are organized according to their chiral dimension,
which counts the number of derivatives and/or meson mass insertions.
Here, we work to second order in the corresponding small parameter $q$
(which is a generic symbol for an external momentum or pion mass). 
Consequently, the effective Lagrangian consists of the following pieces:
\begin{equation}
{\cal L}_{\rm eff} = {\cal L}_{\pi\pi}^{(2)} + {\cal L}_{\pi N}^{(1)}
 + {\cal L}_{\pi N}^{(2)}  \,\, ,
\end{equation}
where the index $(i)$ gives the chiral dimension.
The form of ${\cal L}_{\pi\pi}^{(2)}+ {\cal L}_{\pi N}^{(1)} $ is standard.
Let us discuss in some more detail the terms appearing in 
$ {\cal L}_{\pi N}^{(2)}$. For the reasons outlined before, we treat
the spin--1/2 fields, i.e. the nucleons, relativistically.
The terms of the dimension two effective relativistic pion--nucleon Lagrangian
relevant to our calculation read \cite{gss} \cite{krause}
\beqa \label{lpinrel}
{\cal L}_{\pi N}^{(2)} &=& {c}_1 \bar{\Psi} 
\Psi \Tr(\chi_+) + \dfrac{{c}'_2}{4m} \bar{\Psi}
i \gamma_\mu \stackrel{\leftrightarrow}{D}_\nu \Psi \Tr(u^\mu u^\nu ) -
\dfrac{{c}''_2}{8m^2} \bar{\Psi} 
\stackrel{\leftrightarrow}{D}_\mu \stackrel{\leftrightarrow}{D}_\nu \Psi
\Tr(u^\mu u^\nu ) \nonumber \\
&& \qquad + \, {c}_3 \bar{\Psi} u_\mu u^\mu \Psi
+ {c}_4 \dfrac{i}{4} \bar{\Psi} \sigma_{\mu \nu} [u^\mu, u^\nu ]\Psi
\, \, ,
\eeqa
where $\Psi$ denotes the relativistic nucleon field and with
\beq
\chi_+ = u^\dagger \, \chi \, u^\dagger + u  \, \chi^\dagger \, u \, , \qquad
u_\mu = i \, u^\dagger \, \nabla_\mu \, U \, u^\dagger \,\, .
\eeq
The quantity $\chi$ contains the light quark mass $\hat m$ and external scalar
and pseudoscalar sources \cite{gss} and $\nabla_\mu$ is the standard chiral
covariant derivative acting on the pion fields. The low--energy 
constants ${c}_i$ are normalized such that we can identifiy them with the 
corresponding low--energy constants  of the heavy baryon
Lagrangian (truncated at order $q^2$) 
(for definitions, see e.g. \cite{bkmrev}). Their numerical values have
been determined in \cite{bkmppn} from a tree level fit to pion--nucleon
scattering (sub)threshold parameters,
\beqa \label{cival}
c_1   &=&  -0.64 \pm 0.14 \, \, , \quad 
c_2'  =  -5.63 \pm 0.10 \, \, , \quad 
c_2'' =   7.41 \pm 0.10 \, \, , \nonumber \\
c_3   &=&  -3.90 \pm 0.09 \, \, , \quad 
c_4   =   2.25 \pm 0.09 \, \, ,
\eeqa
with all numbers given in GeV$^{-1}$. It is important to stress that
the effective Lagrangian we use does not incorporate any resonances
like e.g. the $\Delta(1232)$ or the $N^* (1440)$ as dynamical degrees
of freedom. Their contribution is encoded in the numerical values of
the low--energy constants $c_i$, for a more detailed discussion see
\cite{bkmlec}. Clearly, one might doubt that above threshold such a
procedure is sufficient, since for example it does not account for
the effects related to the widths of these resonances. If one expands
the resonance propagators in powers of momentum transfer over
resonance mass, one gets to leading order such dimension
two pion--nucleon couplings as used here (plus an infinite series of higher
order terms).  Our study should therefore
reveal at which kinematics and for which observables explicit
resonance effects can be most easily seen. This would be helpful for
pinning down various resonance couplings which are not too well
constraint by the present models (when global fits are performed).
Finally, we stress that the dimension two operators encode also
information about t-channel meson exchanges, e.g. the constant $c_1$
is essentially saturated by scalar meson (correlated $2\pi$, to be
precise) exchange and $c_4$ receives an important contribution from
the $\rho$ \cite{bkmlec}.

\subsection{Chiral expansion of the invariant amplitudes}

Let ${\cal B}$ be a generic symbol for any one of the $B_i^j$. The chiral
expansion considered here takes the form
\beq
{\cal B} = {\cal B}^{\rm Born} + {\cal B}^{c_i} \,\,\, .
\eeq
Consider first the Born terms with insertions solely from 
${\cal L}_{\pi N}^{(1)}$ shown in Fig.~1. Diagram~A involves
the $NN3\pi$--vertex genuine to $\pi N \to \pi \pi N$ at tree
level. The lowest order four--pion interaction appears in the
pion pole graph (B). The graphs of the classes C, D, E and F
contain one pion--nucleon vertex and one Weinberg $NN\pi\pi$ interaction,
thus these scale as $g_A$. The graphs belonging to the classes G, H and
I each contain three pion--nucleon vertices and thus scale as $g_A^3$.
In Fig.~2, we show the term with exactly one insertion from 
${\cal L}_{\pi N}^{(2)}$, i.e. these are proportional to the low--energy
constants $c_i$. Note that there is no higher order $NN3\pi$--vertex
from the dimension two Lagrangian because of isospin. Note also that the
graphs of type J,~K,~L and M are topologically identical to the ones of
classes N,~O,~P and Q. However, while the operator proportional to $c_4$
contributes to the $B_i^4$, the ones $\sim c_1, c_2', c_2''$ and  $c_3$
can not. The explicit formula for the $B_i^j$ for the various (classes of)
diagrams are given in appendix~B.


\section{Results and discussion}

In this section, we show the results for the total and double
differential cross sections as well as the angular correlation
function in comparison to the available data.  
Our input parameters are $M_\pi =139.57\,$MeV, $m= 938.27\,$MeV,
$F_\pi = 92.4\,$MeV and $g_A = 1.32$ as determined from the
Goldberger--Treiman relation $g_A= F_\pi g_{\pi N}/m$ with $g_{\pi N} = 13.4$.
For the low--energy constants $c_i$, we take the central values given in
Eq.(\ref{cival}). Setting $c_i =0$ corresponds to the tree level (dimension
one)   calculation of Beringer \cite{bering}.

\subsection{Total cross sections}

First, we discuss the data from threshold up to incoming pion kinetic energy
of $T_\pi = 400\,$MeV and then take a closer look at the threshold
regions, $T_\pi \le 210\,$MeV. In all graphs, the solid line gives the result
with the $c_i$ as given in Eq.(\ref{cival}) and the dashed line the one
with the $c_i = 0$.

\noindent \underline {$\pi^+ p \to \pi^+ \pi^+ n$, Fig.~3:} 
The data are from refs. \cite{se91} \cite{ke90} \cite{kr76} \cite{ki62}. 
The lowest order calculation does already 
quite well. Inclusion of the dimension two terms leads to a too large
cross section. This can be traced back to the fact that the Born
term result for the threshold amplitude $D_1$ is closer to the
empirical number than the one with the $c_i$ corrections, compare
Eq.(\ref{D12}).

\noindent \underline {$\pi^+ p \to \pi^+ \pi^0 p$, Fig.~4:} 
There are recent data from \cite{poc} and a
few older data in this kinematical range, see refs. \cite{ba75}
\cite{ar72} \cite{ba63}.  
While the lowest order calculation tends to underestimate the data
above the threshold region, the $c_i$ corrections lead to a somewhat
too large total cross section, in particular when compared only
to the data of \cite{poc}.

\noindent \underline {$\pi^- p \to \pi^+ \pi^- n$, Fig.~5:} 
The data are from refs. \cite{ke89} \cite{bj80} \cite{jo74} \cite{bl70}
\cite{sa70} \cite{ba65} \cite{bl63} \cite{de61} \cite{pe60}. 
The lowest order calculation clearly 
underestimate the cross section even close to threshold.  
Inclusion of the dimension two terms leads to an astonishingly good
description of the total cross section up to $T_\pi = 400\,$MeV.

\noindent \underline {$\pi^- p \to \pi^0 \pi^0 n$, Fig.~6:} 
The data are from refs. \cite{lo91a} \cite{be80} \cite{be78} \cite{bu77}
\cite{kr75}. The lowest order calculation clearly 
underestimates the cross section even close to threshold.  
Inclusion of the dimension two terms leads to a good
description of the total cross section up to $T_\pi = 230\,$MeV and
a slight underestimation for larger pion energies.

\noindent \underline {$\pi^- p \to \pi^0 \pi^- p$, Fig.~7:} 
The data are from refs. \cite{ke89} \cite{jo74} \cite{bl63}. 
Inclusion of the dimension two terms leads to an improved
description of the total cross section up to $T_\pi = 250\,$MeV and
a slight overestimation for larger pion energies whereas the lowest
order result is consistently below the data for $T_\pi > 230\,$MeV.

\noindent \underline {Threshold region:} The threshold regions for
all five channels are shown in Figs.~8 and 9. In all but one case,
i.e. for $\pi^+ p \to \pi^+ \pi^+n$, the inclusion of the terms
$\sim c_i$ clearly improves the description of the data. As already noted
in the introduction, this improvement is not due to the strong pionic
rescattering in the isospin zero, S--wave as originally thought \cite{bkmplb}.
It should be noted, however, that only for $\pi^+ p \to \pi^+ \pi^+ n$ and
$\pi^- p \to \pi^0 \pi^0 n$, {\it precise} data close to threshold exist

\subsection{Double differential cross sections}

Double differential cross sections have been published by Manley
\cite{manl} and by M\"uller et al. \cite{mue} for the reaction
$\pi^- p \to \pi^+ \pi^- n$. In Fig.~10, we show the results in comparison
to the data of \cite{manl} at $\sqrt{s}  = 1.242\,$GeV and for $T_{\pi^+} =
\omega_1- M_\pi =6.5, 10.95$ and $15.35\,$MeV, in order. Fig.~11 comprises the
results for $\sqrt{s}  = 1.262\,$GeV and  $T_{\pi^+}=\omega_1-M_\pi= 10.4,17.5$
and $24.6\,$MeV. In all cases, our full calculation describes well the data
whereas the leading order calculation is systematically below these. However,
we are not able to describe the data of \cite{mue} at $\sqrt{s}  = 1.301\,$GeV
as shown in Fig.~12. The two solid lines in that figure refer to the
angular range of $x_1 = \cos\theta_1$, $0.03 < x_1 < 0.35$. 
The strong peaking of the data around $T_{\pi^+}=\omega_1-M_\pi\simeq 25\,$MeV 
can not be reproduced. Two comments are in order. The experiment of \cite{mue}
was not intended primarily to measure the double differential cross section
but rather the angular correlation function, see the next paragraph. It is,
however, claimed in \cite{mue} that their results for $d^2 \sigma / d\omega_1 d
\Omega_1$  agree with the ones of Manley \cite{manl} \cite{man2} once a scaling
factor to account for the different incident pion kinetic energies is applied
and the different laboratory angles of both experiments are considered.
Other double differential cross sections for
the same process measured by the Erlangen group are given in \cite{boh}
\cite{malz}. These span the range of the incoming pion kinetic energy
$T_{\pi^-}$ from 218 to 330 MeV. The data for the lower energies are well
described within our approach, as shown in Fig.~13. For beam energies
$T_{\pi^-} = 246$ and $330\,$MeV, our results underestimate the data, but not
as strongly as for the data of \cite{mue} at  $T_{\pi^-} = 284\,$MeV.

\subsection{Angular correlation functions}

An extensive study of the angular correlation function $W$ for 
$\pi^- p \to \pi^+ \pi^- n$ at $\sqrt{s} = 1.301\,$GeV is given in 
\cite{mue}. For six values of the momentum of the positive pion,
$|\vec q_1\,| = 60.5, 86, 100, 113, 127$ and $151.5\,$MeV\footnote{We
use the central values of the pertinent momentum bins.} tables are
given for $W$ as a function of $\phi_2$ (with $\theta_2$ fixed) and
of $\theta_2$ (with $\phi_2$ fixed). We scanned through all of these
data and found the following results. For fixed $\theta_2$, the angular 
correlation function is well described as long as $\phi_2 > 50^\circ$.
For smaller azimuthal angles, we consistently underestimate the data,
a typical example is shown in Fig.~14. On the other hand, for the
fixed values of $\phi_2$ given in \cite{mue}, the $\theta_2$ dependence
is satisfactorily described, two examples are given in Fig.~15.
We have also considered the angular correlation functions given in
\cite{boh} and \cite{malz} and found similar trends, i.e. in general
a good description of $W$ as long as the dimension two terms are taken
into account.   

\subsection{Comparison to other calculations}

As already stated, the calculation of \cite{jemi} is in spirit closest
to ours. They use the relativistic pion--nucleon Lagrangian to leading
order\footnote{The are not using heavy baryon CHPT as claimed in their
paper.} and add explicit $\Delta$ and Roper degrees of freedom, but
no heavy meson ($\sigma, \, \rho$) exchanges. Overall, their results are
not significantly better than ours. It is important to note that these
authors did not try a best fit by fine tuning the various coupling
constants. Our conclusion is similar to theirs, namely that for really
testing the constraints from chiral symmetry, one needs more precise
data at low energies. The model of \cite{oset} is more elaborate
concerning the inclusion of meson exchanges and is more tuned to the
resonance  than to the threshold region. Similarly, the work of the Erlangen
group \cite{jaeck} shows the importance of the non--linear pion
couplings at tree level but does not include higher derivative
operators as done here. Our appraoch lends more credit to the
conclusions of these papers that the intricate dynamics of the
reaction $\pi N \to \pi \pi N$ reveals itself best in double
differential cross sections and angular correlation functions but not
in the total cross sections.


\section{Summary and conclusions}

In this paper, we have studied the reaction $\pi N \to \pi \pi N$
in relativistic baryon chiral perturbation theory at tree level 
including the dimension two pion--nucleon Lagrangian. This study was 
motivated by the fact that at threshold
 the dominant corrections to the lowest
order tree graphs indeed come from tree graphs with exactly one
insertion from ${\cal L}_{\pi N}^{(2)}$ and not from loop graphs,
as shown in \cite{bkmppn}. The values of the relevant low--energy
constants can be fixed from a few pion--nucleon scattering data. These
low--energy constants encode information about baryonic and mesonic
resonance excitations, in particular related to the $\Delta$, the
Roper, the $\rho$ and the correlated two--pion exchange in the 
scalar--isoscalar channel \cite{bkmlec}. 
Consequently, all our results concerning single pion production
are based on a truely parameter--free calculation.
The pertinent results of this study can be summarized as follows:

\begin{enumerate}

\item[$\circ$] For pion energies up to $T_\pi = 250\,$MeV, in all but
one case the inclusion of the contribution $\sim c_i$ clearly improves
the description of the total cross sections, most notably in the
threshold region for $\pi^- p \to \pi^0 \pi^0 n$. Up to $T_\pi =
400\,$, the trend of the data can be described although some
discrepancies particularly towards the higher energies persist, see
Figs.~3$\ldots$9. This
indicates that our approach is best suited for the threshold regions.

\item[$\circ$] Double differential cross sections for 
$\pi^- p \to \pi^+ \pi^- n$ at incident pion energies
below $T_\pi = 250\,$MeV are well described (see Figs.~10,11,13).
We are not able to reproduce the strong peak of $d^2 \sigma /d\omega_1 
d\Omega_1$ as a function the outgoing $\pi^+ $ energy  at
$\sqrt{s} = 1.301\,$GeV found in \cite{mue}, compare Fig.~12. 

\item[$\circ$] The angular correlation function for the reaction
$\pi^- p \to \pi^+ \pi^- n$ as given in \cite{mue} can be reproduced
fairly well for all energy bins  and sets of
polar and azimuthal angles given in that paper, with the exception
of small polar angles for the final--state negative pion, compare
Figs.~14 and 15. 

\end{enumerate}

\bigskip

\section*{Acknowledgements}

We thank W. Eyrich and M. Moosburger for valuable information on the
experimental data.

\newpage

\appendix
\def\theequation{\Alph{section}.\arabic{equation}}
\setcounter{equation}{0}
\section{Kinematics and weight functions}

First, we express all possible scalar products of the nucleon and
pion momenta in the initial and final state in terms of the
five Mandelstam invariants, see Eq.(\ref{Mand}), and the particle masses,
\begin{equation} 2p_1\cdot k=s-m^2-M_\pi^2,\quad 2p_1\cdot q_1=s-s_2+t_1-
M_\pi^2, \quad 2p_1\cdot q_2=s-s_1+t_2-M_\pi^2, \end{equation}
\begin{equation} 2p_1\cdot p_2=s_1+s_2-s-t_1-t_2+m^2+M_\pi^2, \quad 2k\cdot q_1
=2M_\pi^2-t_1, \quad 2k\cdot q_2=2M_\pi^2-t_2, \end{equation}
\begin{equation} 2k\cdot p_2=s+t_1+t_2-m^2- 3M_\pi^2, \quad 2q_1\cdot q_2=s-s_1
-s_2+m^2, \end{equation} \begin{equation}2q_1\cdot p_2=s_1-m^2- M_\pi^2, \quad
2q_2\cdot p_2=s_2-m^2-M_\pi^2 \,\, .\end{equation} 

\noindent The weight functions $y_{ij}$ appearing in 
Eqs.(\ref{xstot},\ref{xsdd},\ref{xstd}) are given by:
\begin{eqnarray} y_{11} &=& s_1+s_2-t_1-t_2-s-m^2+M_\pi^2\,,\nonumber \\ y_{12}
&=& m(2(s_1+s_2-s-m^2)-t_1-t_2)\,,\nonumber \\y_{13} &=&m(t_2-t_1)\,,\nonumber 
\\ y_{14} &=& (s_1-s_2)(s-s_1-s_2+m^2)+(t_1-t_2)(m^2+M_\pi^2)+s_1t_2-s_2t_1\,,
\nonumber \\ y_{22} &=& (s_1+s_2)(4m^2-M_\pi^2)+(t_1+t_2)(s-m^2)+s^2-3s(2m^2+
M_\pi^2) \nonumber \\ & &-3m^2(m^2+M_\pi^2)+2M_\pi^4\,, \nonumber \\ y_{23} &=&
(s_1-s_2)(s-m^2-2 M_\pi^2) +(t_2-t_1)(m^2+M_\pi^2)+s_1t_1-s_2t_2\,,\nonumber \\
y_{24} &=& m (2s_1- 2s_2+t_1-t_2)(s-s_1-s_2+m^2+2M_\pi^2)\,,\nonumber \\ y_{33}
&=& (s_1+s_2)(2s-2 m^2- 3M_\pi^2)+(t_1+t_2)(2M_\pi^2-m^2-s) -4s_1s_2 \nonumber
\\ & & +2s_1t_1+2 s_2t_2-s^2 +s(2m^2+3M_\pi^2) +3m^4-5 m^2M_\pi^2-2M_\pi^4 \,,
\nonumber \\ y_{34} &=& m(2s +t_1+t_2-2m^2-4M_\pi^2)(s-s_1-s_2+m^2-2M_\pi^2)\,,
\nonumber \\y_{44} &=& (s-s_1- s_2+m^2)\bigl[(s-m^2)(s+t_1+t_2-m^2)+(s_1-s_2)
(s_2-s_1+t_2-t_1)\bigr] \nonumber \\ & & +M_\pi^2 \bigl\{4M_\pi^2(2s-M_\pi^2)
+m^2\bigl[ 3m^2-8M_\pi^2 +6(s-s_1-s_2)+2(t_1+t_2)\bigr] \nonumber \\ & &-5s^2+
2s( s_1+s_2-t_1-t_2)-s_1^2-s_2^2+6 s_1s_2 +2(s_1-s_2)(t_2-t_1)\bigr\}\,.
\end{eqnarray} 
and they are symmetric under interchange of the indices,  $y_{ij}=y_{ji}$. 


\setcounter{equation}{0}
\section{Invariant amplitudes $B_i^j$}
\label{bij}

In this appendix we give the
invariant amplitudes $B_i^j$ for individual (classes of)
diagrams. The $B_i^j$ not specified are zero for the class of diagrams just
considered. 
The following abbreviations for kinematical quantities are used: 
\begin{equation} d_1=s_1-m^2\,,\quad d_2=s_2-m^2\,,\quad
d_3=2M_\pi^2-s+s_1-t_2\,,  \nonumber \end{equation}
\begin{equation} d_4=2M_\pi^2-s+s_2-t_1\,, \quad  d_5=m^2+4M_\pi^2-s-t_1-t_2 
\,, \quad d_6=s-m^2 \,\, . \nonumber \end{equation} 

\medskip

\noindent\underline{Diagram A:}
\begin{equation} B_2^1=2\,, \quad B^2_1=4m\,, \quad B_2^2=B_3^3=-1\,. \nonumber
\end{equation} 

\noindent\underline{Diagram B:}
\begin{equation} B_1^1={8m(d_1+d_2-d_6-M_\pi^2)\over 2M_\pi^2-d_1-d_2-d_5} \,, 
\quad  B_1^2={4m(2M_\pi^2-t_1-t_2)\over 2M_\pi^2-d_1-d_2-d_5} \,, 
\quad  B_1^3={4m(t_1-t_2)\over 2M_\pi^2-d_1-d_2-d_5}\,.\nonumber\end{equation} 
Here, the separation of the invariant amplitudes between diagram A and B
results from using the $\sigma$-model gauge, $U= \sqrt{1-\vec
  \pi^2/F_\pi^2} + i \vec \tau \cdot \vec \pi/F_\pi$. 
We have checked that other parametrizations lead
to the same result for diagram A plus B. 

\noindent\underline{Diagram C:}
\begin{equation} B^3_1=B^4_1 =2m(s_2-s_1)d_6^{-1}\,, \quad B^3_3=B^4_3 =1\,, 
\quad B^3_4=B^4_4=-2md_6^{-1}\,. \nonumber \end{equation}

\noindent\underline{Diagram D:}
\begin{equation} B^3_1=-B^4_1 =2m(d_3-d_4)d_5^{-1}\,, \quad B^3_3=-B^4_3 =1\,, 
\quad B^3_4=-B^4_4=2md_5^{-1}\,. \nonumber \end{equation}

\noindent\underline{Diagrams E:}
\begin{equation} B^1_1=-2B^2_1 =2m(d_1+d_2-d_6)(d_1^{-1}+d_2^{-1}) \,, \quad 
B^1_2=-2B^2_2 =-3\,, \nonumber \end{equation} \begin{equation}
B^1_4=-2B^2_4=2m(d_2^{-1}-d_1^{-1})\,, \quad B^4_1=2B^3_1
=2m(d_1+d_2-d_6)(d_1^{-1}-d_2^{-1}) \,,\nonumber\end{equation}\begin{equation} 
B^4_3=2B^3_3 =1\,,  \quad B^4_4=2B^3_4=-2m(d_1^{-1}+d_2^{-1})\,. \nonumber 
\end{equation} 

\noindent\underline{Diagrams F:}
\begin{equation} B^1_1=-2B^2_1 =2m(d_1+d_2-d_6)(d_3^{-1}+d_4^{-1}) \,, \quad 
B^1_2=-2B^2_2 =-3\,, \end{equation} \begin{equation}   B^1_4=-2B^2_4=2m(
d_4^{-1}-d_3^{-1})\,, \quad B^4_1=-2B^3_1 =2m(d_1+d_2-d_6)(d_3^{-1}-d_4^{-1})
 \,, \end{equation} \begin{equation}   B^4_3=-2B^3_3 =-1\,,  \quad
 B^4_4=-2B^3_4=-2m(d_3^{-1}+d_4^{-1})\,.\end{equation}  

\noindent\underline{Diagrams G:}
\begin{equation} B^1_1=g_A^2md_6^{-1}\{(s+3m^2)[(s-s_1)d_2^{-1}+(s-s_2)
d_1^{-1}] -4(s+m^2)\}   \,, \end{equation} 
\begin{equation}   B^3_1=B^4_1 =g_A^2m  d_6^{-1}\{(s+3m^2)[(s-s_2)
d_1^{-1}-(s-s_1)d_2^{-1}]+2(s_1-s_2)\} \,, \end{equation} 
\begin{equation} B^1_2=g_A^2 [8m^2d_6^{-1}+1-2m^2(d_1^{-1}+d_2^{-1})] 
\,, \quad  B^3_2=B^4_2 =B^1_3= 2g_A^2m^2( d_2^{-1}-d_1^{-1})\,, \end{equation}
 \begin{equation} B^3_3= B^4_3=g_A^2[-1-2m^2(d_1^{-1}+d_2^{-1})]
 \,, \quad B^1_4=g_A^2m(s+3m^2) d_6^{-1}(d_1^{-1}-d_2^{-1}) \,, \end{equation}
\begin{equation} B^3_4=B^4_4 =g_A^2md_6^{-1}[2+(s+3m^2)(d_1^{-1}+d_2^{-1})]
\,.\end{equation} 

\noindent\underline{Diagrams H:}
\begin{eqnarray}  B^1_1=-B^2_1 &=&  g_A^2m\{2+(s-s_2)d_1^{-1}+
  (s-s_1)d_2^{-1} \nonumber \\ & &+(d_6-d_1-d_2)[(3m^2+s_1)d_1^{-1}d_3^{-1}+
(3m^2+s_2)d_2^{-1} d_4^{-1}]\}\,, \end{eqnarray} 
\begin{equation}   B^4_1 =g_A^2m \{(s-s_2)d_1^{-1}-(s-s_1)d_2^{-1}+(d_6-d_1-
d_2)[(3m^2+s_1)d_1^{-1}d_3^{-1}-(3m^2+s_2)d_2^{-1}d_4^{-1}\}\,, \end{equation} 
\begin{equation} B^1_2=-B^2_2=2g_A^2[1+m^2(d_1^{-1}+d_2^{-1}+d_3^{-1}+d_4^{-1})
] \,, \quad  B^4_2 = 2g_A^2m^2(d_1^{-1}- d_2^{-1}+d_3^{-1}-d_4^{-1})\,, 
\end{equation}
 \begin{equation} B^1_3= -B^2_3=2g_A^2m^2(d_1^{-1}-d_2^{-1}-d_3^{-1}+d_4^{-1})
 \,, \end{equation}  \begin{equation} B^4_3=2g_A^2m^2(d_1^{-1}+d_2^{-1}-
d_3^{-1}-d_4^{-1}) \,, \end{equation}
 \begin{equation} B^1_4=-B_4^2=g_A^2m[(4m^2+d_1+d_3)d_1^{-1}
d_3^{-1}  -(4m^2+d_2+d_4)d_2^{-1}d_4^{-1}]   \,, \end{equation}
\begin{equation} B^4_4 =g_A^2m[(4m^2+d_1+d_3)d_1^{-1}d_3^{-1}
 +(4m^2+d_2+d_4)d_2^{-1}d_4^{-1}]\,. \end{equation} 

\noindent\underline{Diagrams I:}
\begin{equation} B^1_1=g_A^2md_5^{-1}\{(4m^2+d_5)[(d_5-d_4)d_3^{-1}+
(d_5-d_3)d_4^{-1}] -4(d_5+2m^2)\}   \,, \end{equation} 
\begin{equation} B^3_1 = -B^4_1=g_A^2md_5^{-1}\{(d_5+4m^2)[(d_5-d_3)d_4^{-1}-
(d_5-d_4)d_3^{-1}]+2(d_4-d_3)\}\,, \end{equation}  
\begin{equation} B^1_2=g_A^2 [1+8m^2d_5^{-1}-2m^2(d_3^{-1}+d_4^{-1})] 
\,, \quad  B^3_2=-B^4_2 =B^1_3= 2g_A^2m^2( d_3^{-1}-d_4^{-1})\,, \end{equation}
 \begin{equation} B^3_3= -B^4_3=g_A^2[-1-2m^2(d_3^{-1}+d_4^{-1})]
 \,, \quad B^1_4=g_A^2m(d_5+4m^2)d_5^{-1}(d_3^{-1}-d_4^{-1}) \,, \end{equation}
\begin{equation} B^3_4=-B^4_4 =g_A^2md_5^{-1}[-2-(d_5+4m^2)(d_3^{-1}+d_4^{-1})]
\,. \end{equation}

\noindent\underline{Diagrams J:}
\begin{equation} B^1_1=d_6^{-1}\{4(3m^2+s)[4c_1M_\pi^2+c_3(d_6-d_2-d_1)] +
[d_6^2-(d_1-d_2)^2] (2c_2'+c_2''(d_6/(2m^2)+2)] \}   \,, \end{equation} 
\begin{equation}   B^1_2 = -d_6(c_2'+c_2'')/m -md_6^{-1}[32c_1M_\pi^2-c_2''
(d_1-d_2)^2/m^2+8 c_3(d_6-d_2-d_1)] \,, \end{equation} 
\begin{equation} B^1_3=c_2'(s_1-s_2)/m\,, \quad  B^1_4= 2c_2' (s_2-s_1)d_6^{-1}
\,. \end{equation}

\noindent\underline{Diagrams K:}
\begin{equation} B^1_1=d_5^{-1}\{4(4m^2+d_5)[4c_1M_\pi^2+c_3(d_6-d_2-d_1)] +
[d_5^2-(d_3-d_4)^2] (2c_2'+c_2''(d_5/(2m^2)+2)] \}   \,, \end{equation} 
\begin{equation}   B^1_2 = -d_5(c_2'+c_2'')/m -md_5^{-1}[32c_1M_\pi^2-c_2''
(d_3-d_4)^2/m^2+8 c_3(d_6-d_2-d_1)] \,, \end{equation} 
\begin{equation} B^1_3=c_2'(d_4-d_3)/m\,, \quad  B^1_4= 2c_2' (d_4-d_3)d_5^{-1}
\,. \end{equation}

\noindent\underline{Diagrams L:}
\begin{eqnarray}B^2_1&=&16c_1M_\pi^2-2c_3(d_5+d_6)+c_2'(d_6-d_1-d_2)[(d_3-d_6)
d_1^{-1}+(d_4-d_6)d_2^{-1}]     \nonumber \\ & &+c_2''/(2m^2)[(d_6-
M_\pi^2+ t_1/2)(d_4-s+s_2)+(d_6-M_\pi^2+t_2/2)(d_3-s+s_1)] \,,
\nonumber \\ & & 
\end{eqnarray} \begin{eqnarray}B^3_1 &=& 2c_3(t_2-t_1)+c_2'(d_6-d_1-d_2)\{ [
2(M_\pi^2-d_6) -t_2]d_1^{-1} - [2(M_\pi^2-d_6)-t_1]d_2^{-1} \} \nonumber \\
& & +c_2''/m^2 [ m^2(s_2-s_1+(t_2-t_1)/2) +(s-M_\pi^2)(d_3-d_4) \nonumber \\ 
& & +t_2/2(s_1- t_2/2)-t_1/2(s_2-t_1/2)] \,, \end{eqnarray} 
 \begin{eqnarray}   B^2_2 &=& 8c_1 M_\pi^2m(d_1^{-1}+d_2^{-1}) +2c_3m[(t_2-2
M_\pi^2)d_1^{-1}+(t_1-2M_\pi^2)d_2^{-1}]\nonumber\\ & &+c_2'/m[m^2+3M_\pi^2 -
3s +s_1+s_2-3/4(t_1+t_2)]+ c_2''/m [ 2(d_6-M_\pi^2) \nonumber\\
& & + (t_1+t_2)/2  -(s+t_2/2-m^2-M_\pi^2)^2d_1^{-1}-(s+t_1/2-m^2-M_\pi^2)^2
d_2^{-1}] \,, \nonumber \\ & & \end{eqnarray}  
 \begin{eqnarray}   B^3_2 &=& 8c_1 M_\pi^2m(d_1^{-1}-d_2^{-1}) +2c_3m[(t_2-2
M_\pi^2)d_1^{-1}- (t_1-2M_\pi^2)d_2^{-1}]\nonumber\\ & &+c_2'/m[s_1-s_2+3/4(t_1
-t_2)]   +c_2''/m [ (t_2-t_1)/2\nonumber\\ & &-(s+t_2/2-m^2-M_\pi^2)^2d_1^{-1}
+(s+t_1/2-m^2- M_\pi^2)^2d_2^{-1}] \,, \end{eqnarray}  \begin{equation} 
B^2_3 = B^3_2+c_2'(d_4-d_3)/m\,, \end{equation}
\begin{equation} B^3_3= B^2_2+c_2'(3s-m^2-s_1-s_2-d_5)/m\,, \end{equation}
\begin{equation} B^2_4 = c_2'[ (2(M_\pi^2-d_6)-t_2)d_1^{-1}- (2(M_\pi^2-d_6)-
t_1)d_2^{-1}] \,, \end{equation} \begin{equation} B^3_4 = c_2'[2+ (2(M_\pi^2-
d_6)-t_2)d_1^{-1}+ (2(M_\pi^2 -d_6)- t_1)d_2^{-1} \,. \end{equation}

\noindent\underline{Diagrams M:}
\begin{eqnarray}B^2_1&=&16c_1M_\pi^2-2c_3(d_5+d_6)+c_2'(d_6-d_1-d_2)[(d_1-d_5)
d_3^{-1}+(d_2-d_5)d_4^{-1}] \}    \nonumber \\ & &+c_2''/m^2[(d_6-
3M_\pi^2+ t_1/2+t_2)(M_\pi^2-d_1-t_1/2) \, \nonumber \\ & &+(d_6-3M_\pi^2+t_2/2
+t_1)(M_\pi^2-d_2- t_2/2)] \,, 
\end{eqnarray} \begin{eqnarray}B^3_1 &=& 2c_3(t_2-t_1)+c_2'(d_6-d_1-d_2)\{ [
(d_2-d_5)d_4^{-1} - (d_1-d_5)d_3^{-1} \} \nonumber \\
& & +c_2''/m^2 [ (d_6-3M_\pi^2)(s_1-s_2)+(t_1-t_2)(s/2-M_\pi^2) \nonumber \\ 
& & +(t_1/2+t_2)(t_1/2+s_1))-(t_2/2+t_1)(t_2/2+s_2)] \,, \end{eqnarray} 
 \begin{eqnarray}   B^2_2 &=& 8c_1 M_\pi^2m(d_3^{-1}+d_4^{-1}) +2c_3m[(t_2-2
M_\pi^2)d_4^{-1}+(t_1-2M_\pi^2)d_3^{-1}]\nonumber\\ & &+c_2'/m[s-3m^2-5M_\pi^2
+s_1+s_2+5/4(t_1+t_2)]  + c_2''/m [ d_1+d_2+(t_1+t_2)/2 \nonumber\\ &
&-2M_\pi^2 -(s_1+t_1/2-m^2-M_\pi^2)^2d_3^{-1}-(s_2+t_2/2-m^2-M_\pi^2)^2
d_4^{-1}] \,, \end{eqnarray}  
 \begin{eqnarray}   B^3_2 &=& 8c_1 M_\pi^2m(d_4^{-1}-d_3^{-1}) +2c_3m[(t_2-2
M_\pi^2)d_4^{-1}- (t_1-2M_\pi^2)d_3^{-1}]\nonumber \\ & &+c_2'/m[s_2-s_1+(t_2
-t_1)/4]  +c_2''/m [ s_2-s_1+(t_2-t_1)/2\nonumber\\ & &+(s_1+t_1/2-m^2-
M_\pi^2)^2d_3^{-1} -(s_2+t_2/2-m^2- M_\pi^2)^2d_4^{-1}] \,, \end{eqnarray}  
\begin{equation}  B^2_3 = B^3_2+c_2'(s_1-s_2)/m\,, \end{equation}
\begin{equation}  B^3_3 = B^2_2+c_2'(2d_5-d_1-d_2)\,, \end{equation}
\begin{equation} B^2_4 = c_2'[ (2(d_1-M_\pi^2)+t_1)d_3^{-1}- (2(d_2-M_\pi^2)+
t_2)d_4^{-1}] \,, \end{equation} \begin{equation} B^3_4 = c_2'[2- (2(d_1-
M_\pi^2)+t_1)d_3^{-1}- (2(d_2-M_\pi^2)+ t_2)d_4^{-1}] \,. \end{equation}

\noindent\underline{Diagram N:}
\begin{equation} B^3_2=B^4_2=4c_4m(s_2-s_1)d_6^{-1}\,, \quad
B^3_3=B^4_3=4c_4m\,, \quad B^3_4=B^4_4=-2c_4(s+3m^2)d_6^{-1}\,. \end{equation} 

\noindent\underline{Diagram O:}
\begin{equation} B^3_2=-B^4_2=4c_4m(d_3-d_4)d_5^{-1}\,, \quad
B^3_3=-B^4_3=4c_4m\,,\quad B^3_4=-B^4_4=2c_4(d_5+4m^2)d_5^{-1}\,.\end{equation}

\noindent\underline{Diagram P:}
\begin{equation} B^1_1=-2B^2_1=2c_4\{s+7m^2-d_5+4m^2[(s_2-s)d_1^{-1}+(s_1-s)
d_2^{-1} ] \}\,, \end{equation}
\begin{equation} B^4_1=2B^3_1=2c_4\{2(s_2-s_1)+t_2-t_1+4m^2[(s_2-s)d_1^{-1}-
(s_1-s) d_2^{-1} ] \}\,, \end{equation}
\begin{equation} B^1_2=-2B^2_2=2c_4m[-8+(2(d_6-M_\pi^2)+t_2)d_1^{-1}+(2(d_6-
M_\pi^2)+t_1) d_2^{-1} ] \} \,,\end{equation}
\begin{equation} B^4_2=2B^3_2=2c_4m[(2(d_6-M_\pi^2)+t_2)d_1^{-1}-(2(d_6-
M_\pi^2)+t_1) d_2^{-1} ] \} \,,\end{equation}
\begin{equation} B^1_3=-2B^2_3=B^4_2 \,, \quad B^4_3=2B^3_3 = B^1_2+16 c_4m \,,
\end{equation}
\begin{equation} B^1_4=-2B^2_4=8c_4m^2(d_2^{-1}-d_1^{-1}) \,, \quad B^4_4=2
B^3_4=-4c_4[1+2m^2(d_1^{-1}+d_2^{-1})  ] \,.\end{equation}

\noindent\underline{Diagram Q:}
\begin{equation} B^1_1=-2B^2_1=2c_4[d_5-d_6+4m^2(d_1+d_2-d_6)(d_3^{-1}+
d_4^{-1} )] \,, \end{equation}
\begin{equation} B^4_1=-2B^3_1=2c_4[2(s_2-s_1)+t_2-t_1+4m^2(d_1+d_2-d_6)
 (d_3^{-1}- d_4^{-1}) ] \,, \end{equation}
\begin{equation} B^1_2=-2B^2_2=4c_4m[-4+(m^2+3M_\pi^2-s-t_2-t_1/2)d_3^{-1}+
(m^2+3M_\pi^2-s-t_1-t_2/2)d_4^{-1} ]\,, \end{equation}
\begin{equation} B^4_2=-2B^3_2=4c_4m[(m^2+3M_\pi^2-s-t_2-t_1/2)d_3^{-1}-(m^2+3
M_\pi^2-s-t_1-t_2/2) d_4^{-1} ] \} \,, \end{equation}
\begin{equation} B^1_3=-2B^2_3=-B^4_2 \,, \quad B^4_3=-2B^3_3 = -B^1_2-16 c_4m
\,, \end{equation}
\begin{equation} B^1_4=-2B^2_4=8c_4m^2(d_4^{-1}-d_3^{-1}) \,, \quad B^4_4=-2
B^3_4=-4c_4[1+2m^2(d_3^{-1}+d_4^{-1})  ]\,. \end{equation}

\vfill\eject


\pagebreak


$\,$

\section*{Figures}

\vspace{1.5cm}
 
\begin{figure}[h]
   \vspace{0.5cm}
   \epsfysize=13cm
   \centerline{\epsffile{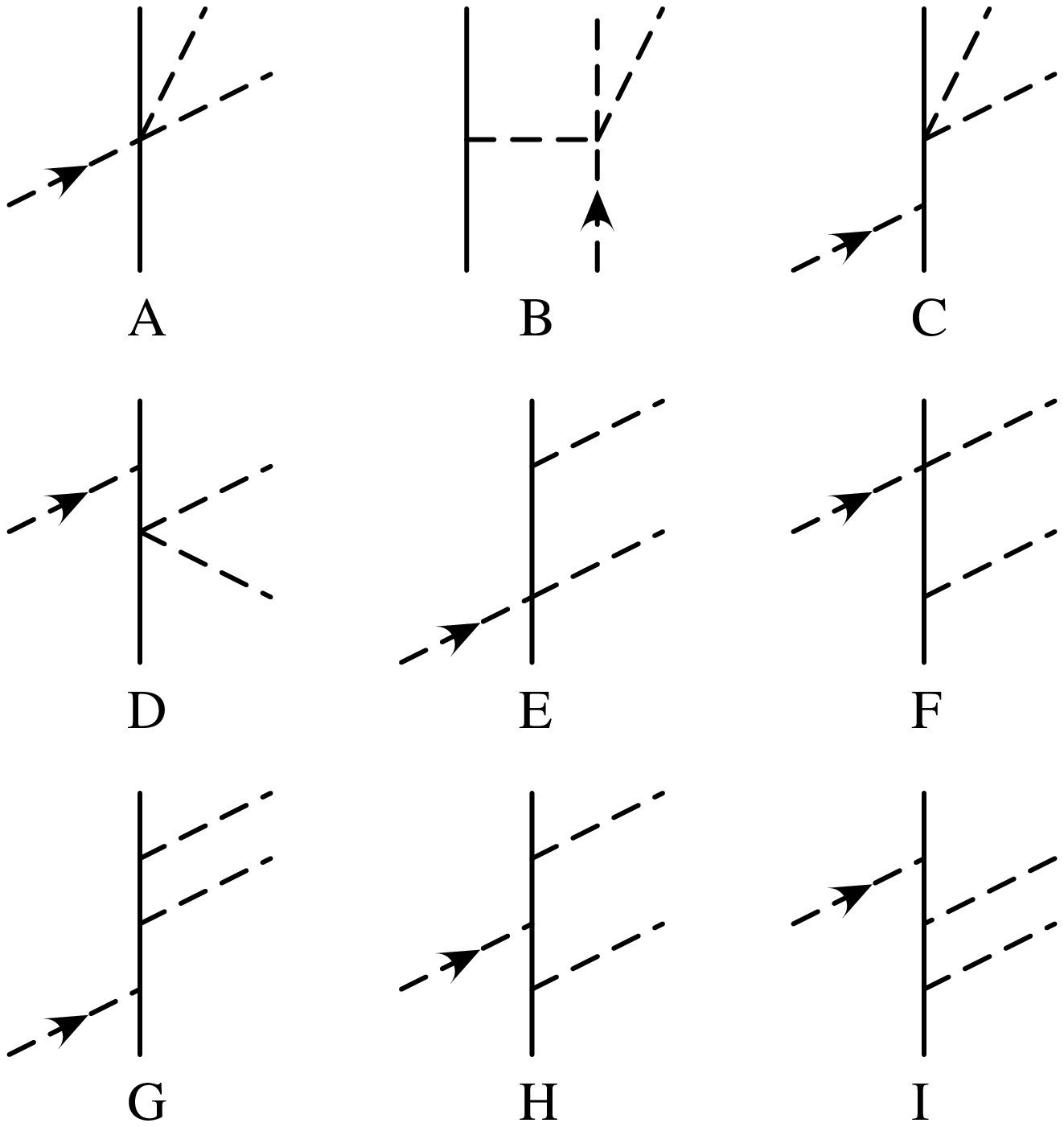}}
   \vspace{1cm}
   \centerline{\parbox{13cm}{\caption{\label{figd1}
Tree graphs with insertions from the dimension one Lagrangian. 
The incoming pion is marked with an arrow.
Diagrams with the final two pions exchanged ($b  \leftrightarrow c$)
are not shown.
  }}}
\end{figure}

\begin{figure}[h]
   \vspace{0.5cm}
   \epsfysize=13cm
   \centerline{\epsffile{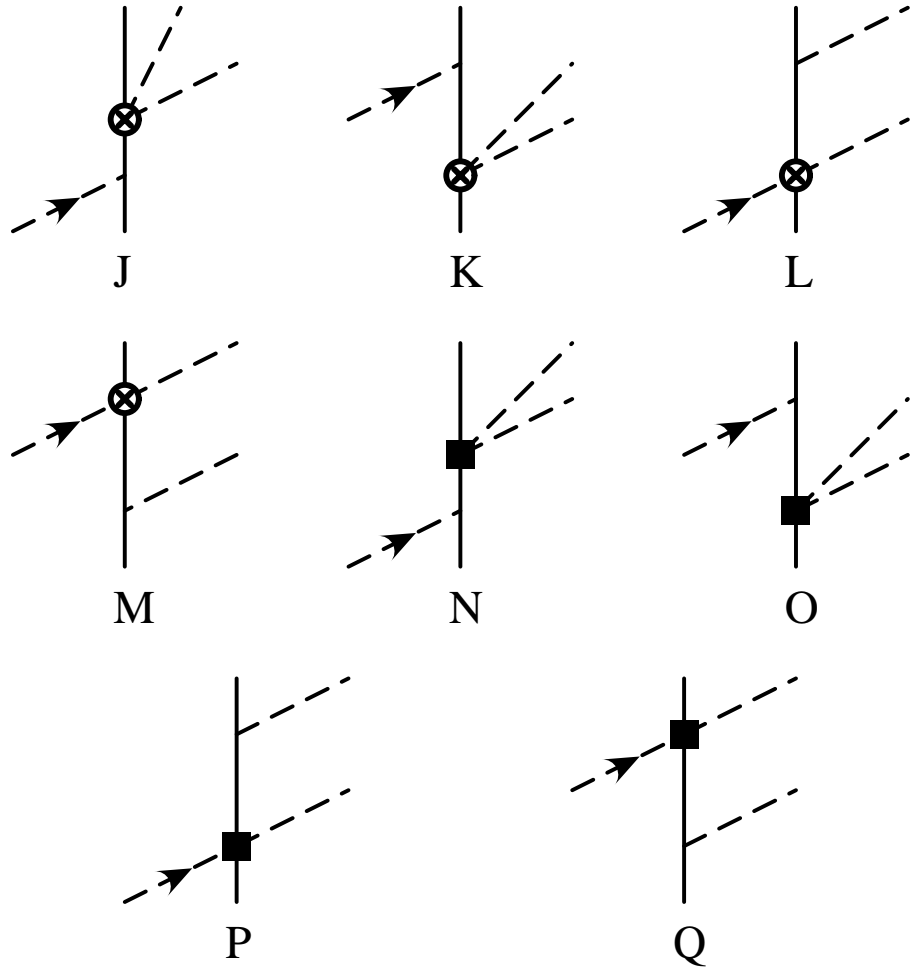}}
   \vspace{1cm}
   \centerline{\parbox{13cm}{\caption{\label{figd2}
Tree graphs with exactly one insertion from the dimension two Lagrangian.
The incoming pion is marked with an arrow.
The circlecross and the box denotes an insertion proporional to $c_1, c_2',
c_2'',c_3$ and $c_4$, respectively.
Diagrams with the final two pions exchanged ($b  \leftrightarrow c$)
are not shown.
  }}}
\end{figure}


\begin{figure}[h]
   \vspace{0.5cm}
   \epsfysize=7.5cm
   \centerline{\epsffile{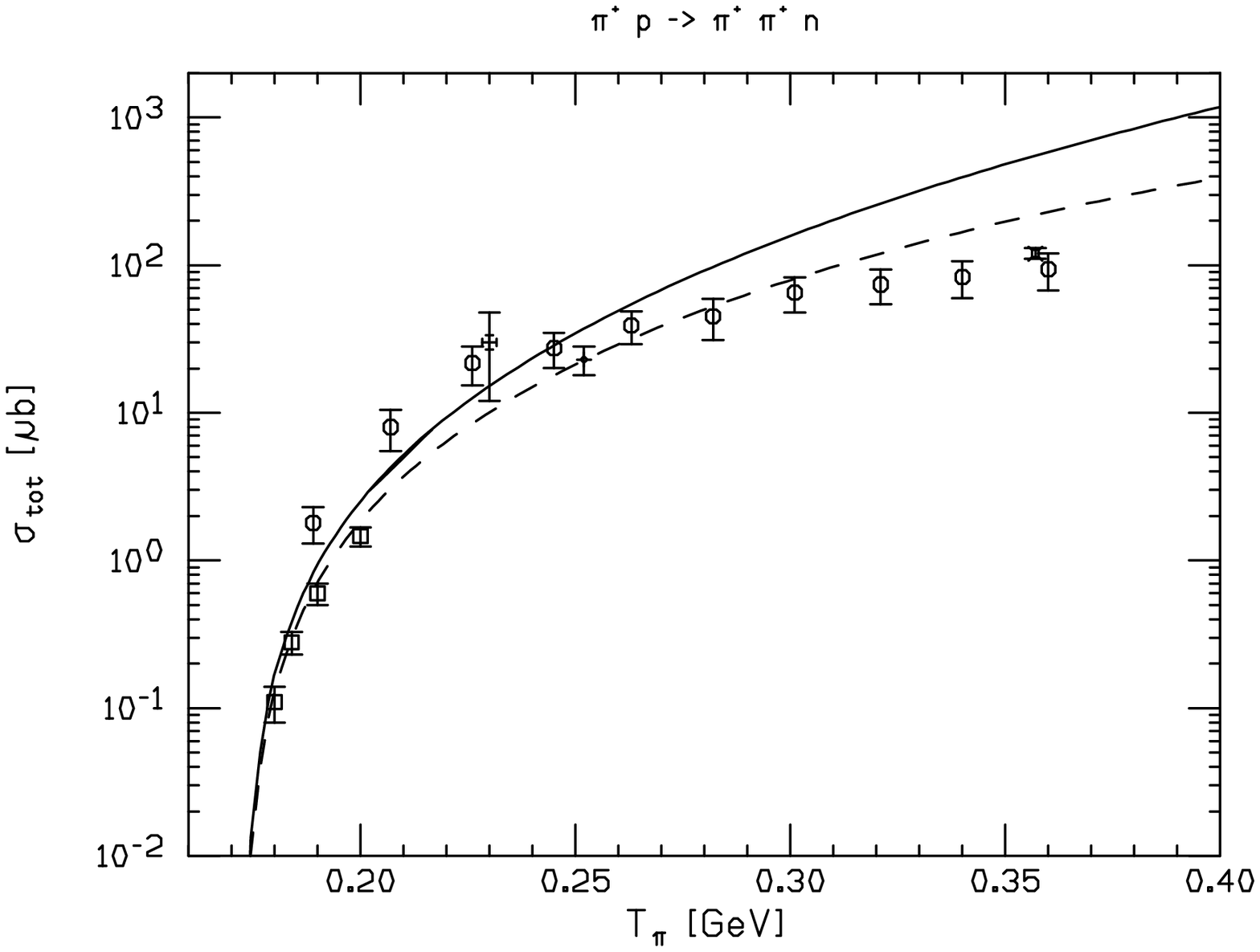}}
   \centerline{\parbox{11cm}{\caption{\label{fig1}
Total cross section for $\pi^+ p \to \pi^+ \pi^+ n$.
  }}}
\end{figure}

\begin{figure}[h]
   \vspace{0.5cm}
   \epsfysize=7.5cm
   \centerline{\epsffile{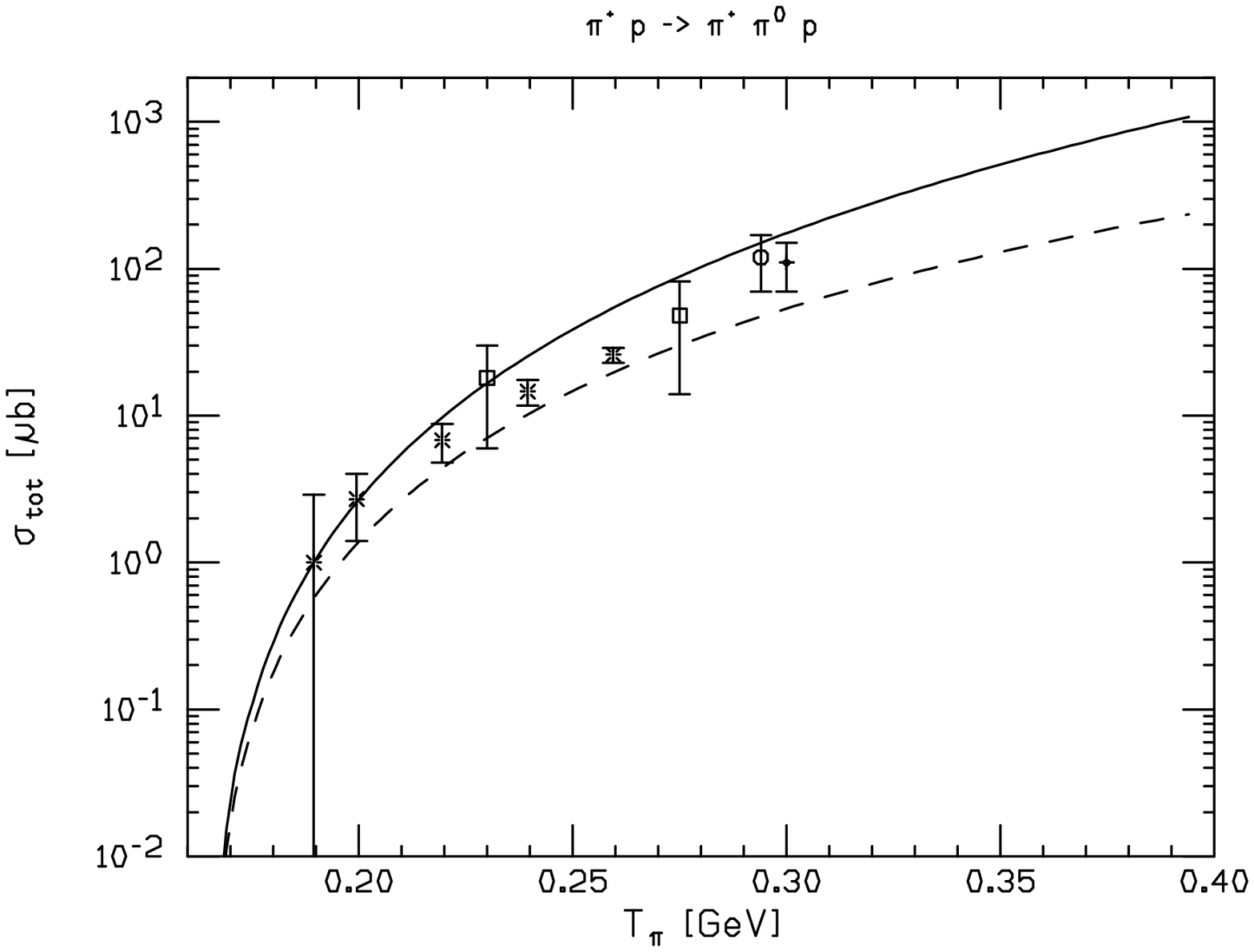}}
   \centerline{\parbox{11cm}{\caption{\label{fig2}
Total cross section for $\pi^+ p \to \pi^+ \pi^0 p$.
  }}}
\end{figure}

\begin{figure}[t]
   \vspace{0.5cm}
   \epsfysize=7.5cm
   \centerline{\epsffile{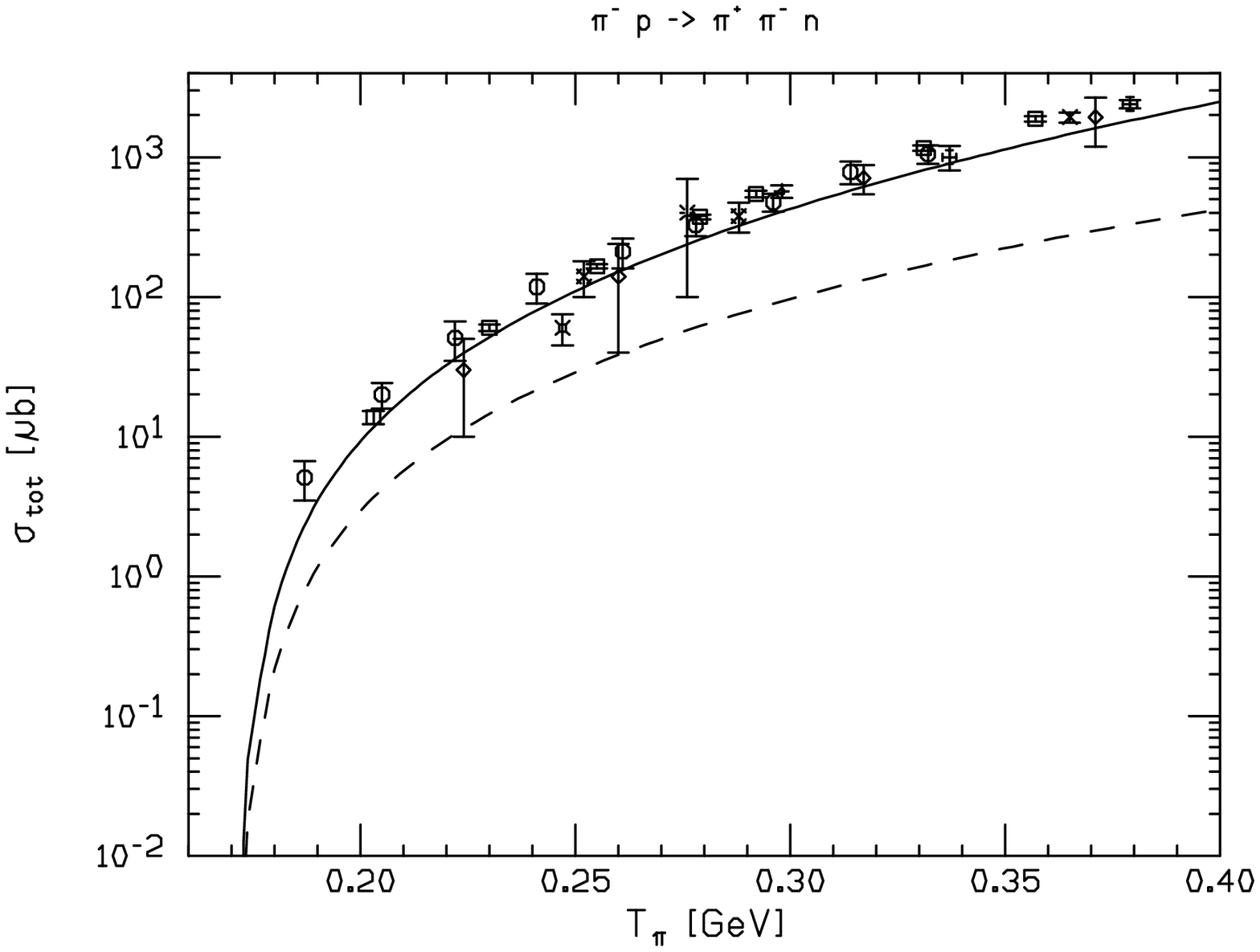}}
   \centerline{\parbox{11cm}{\caption{\label{fig3}
Total cross section for $\pi^- p \to \pi^+ \pi^- n$.
  }}}
\end{figure}

\begin{figure}[b]
   \vspace{0.5cm}
   \epsfysize=7.5cm
   \centerline{\epsffile{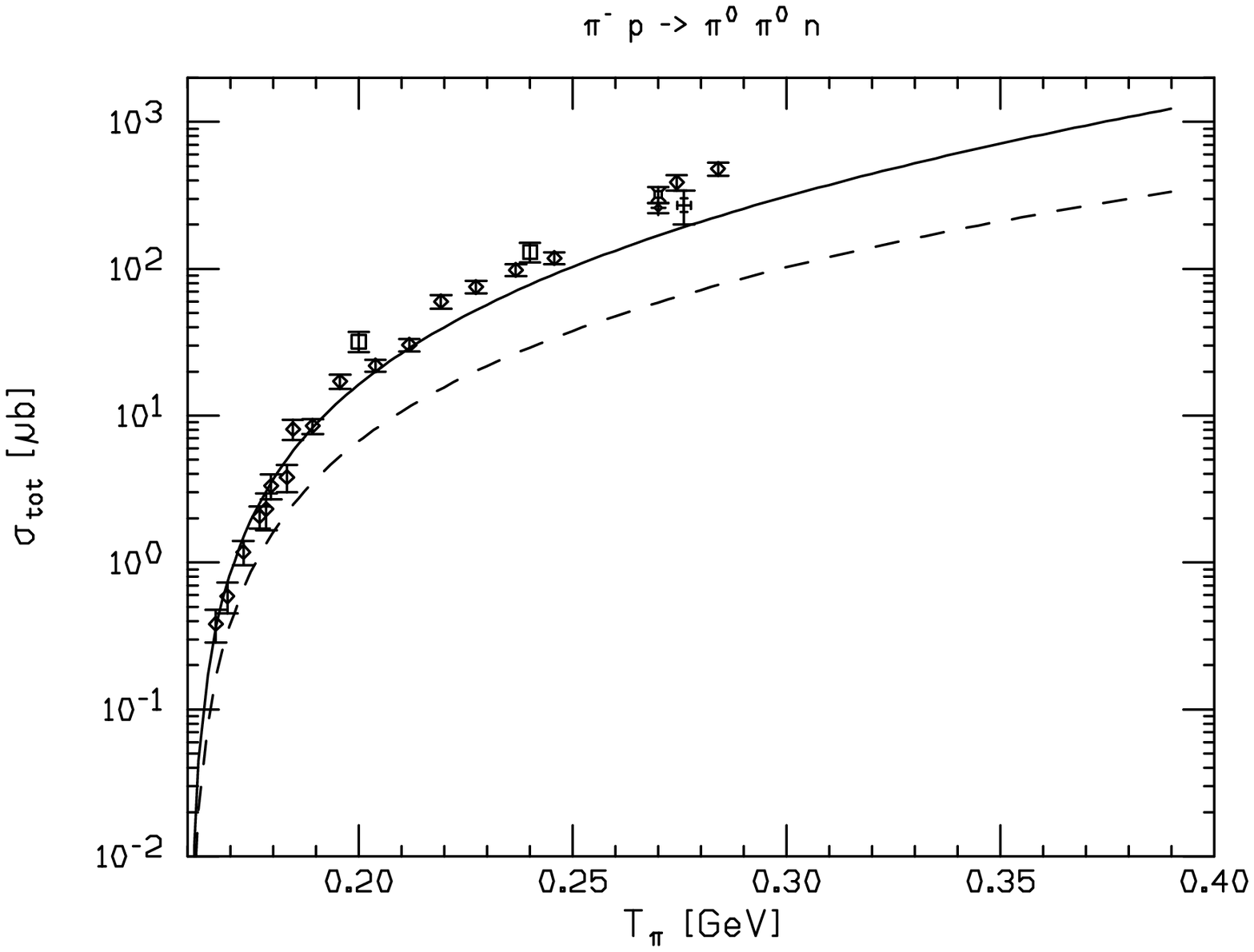}}
   \centerline{\parbox{11cm}{\caption{\label{fig4}
Total cross section for $\pi^- p \to \pi^0 \pi^0 n$.
  }}}
\end{figure}

\begin{figure}[t]
   \vspace{0.5cm}
   \epsfysize=7.5cm
   \centerline{\epsffile{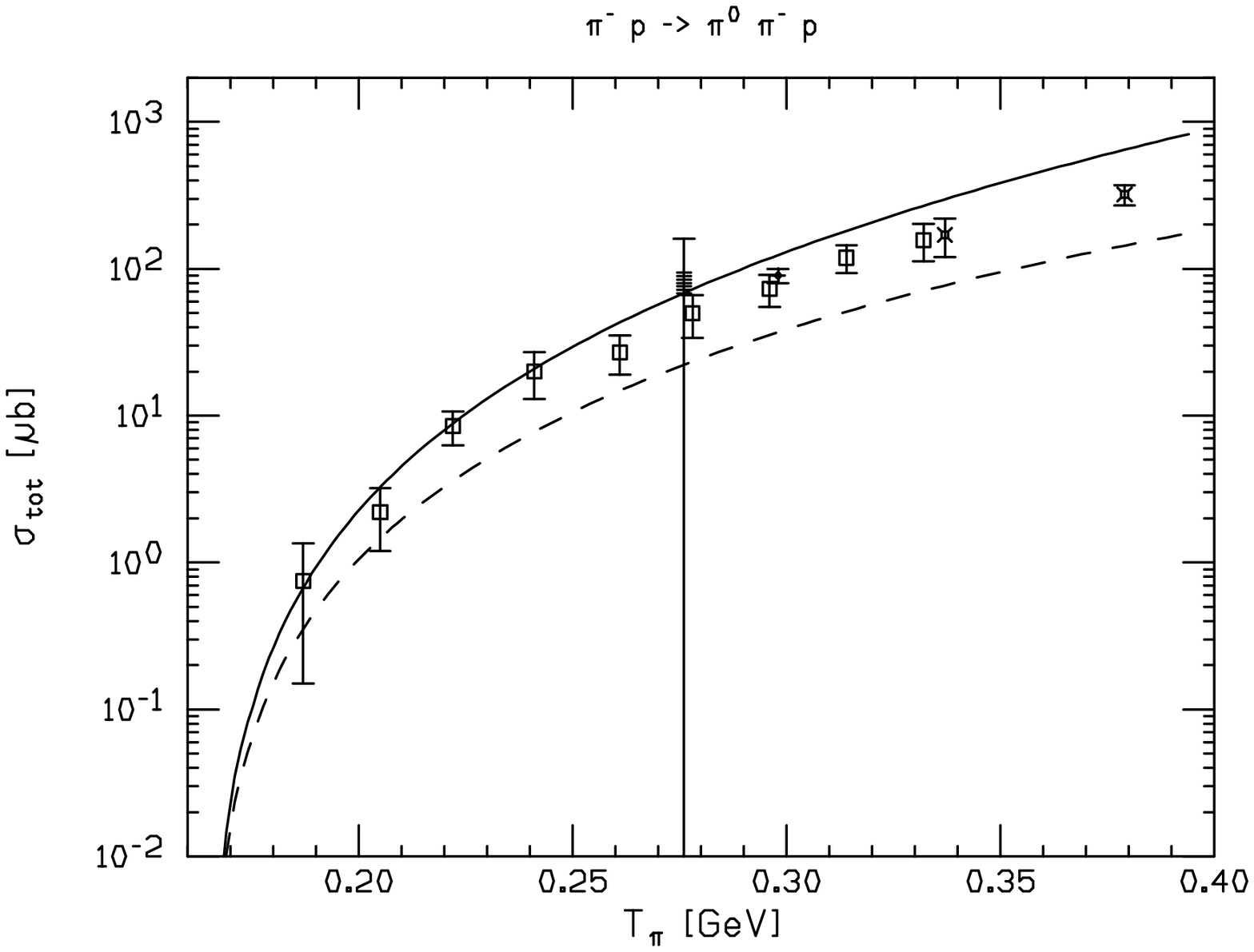}}
   \centerline{\parbox{11cm}{\caption{\label{fig5}
Total cross section for $\pi^- p \to \pi^0 \pi^- p$.
  }}}
\end{figure}

\begin{figure}[b]
   \vspace{0.5cm}
   \epsfxsize=7.5cm
   \centerline{\epsffile{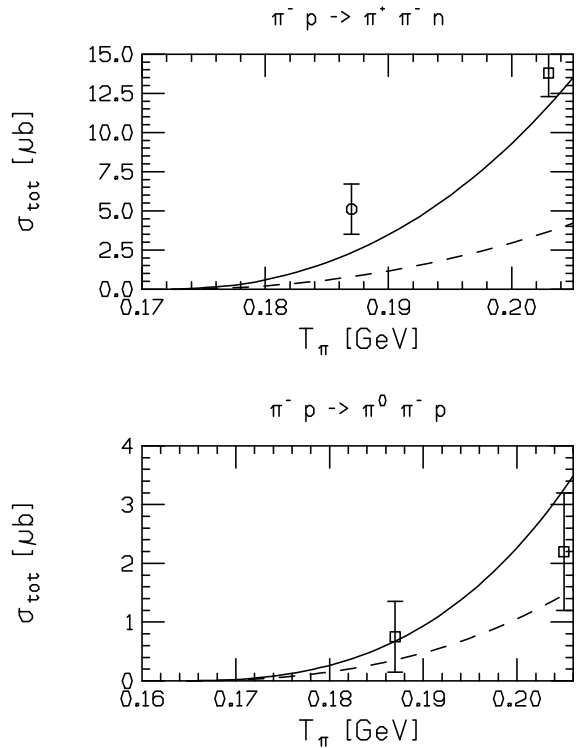}}
   \centerline{\parbox{11cm}{\caption{\label{fig6}
Threshold cross section for $\pi^- p \to \pi^+ \pi^- n$ and 
$\pi^- p \to \pi^0 \pi^- p$.
  }}}
\end{figure}

\begin{figure}[t]
   \vspace{0.5cm}
   \epsfxsize=7.5cm
   \centerline{\epsffile{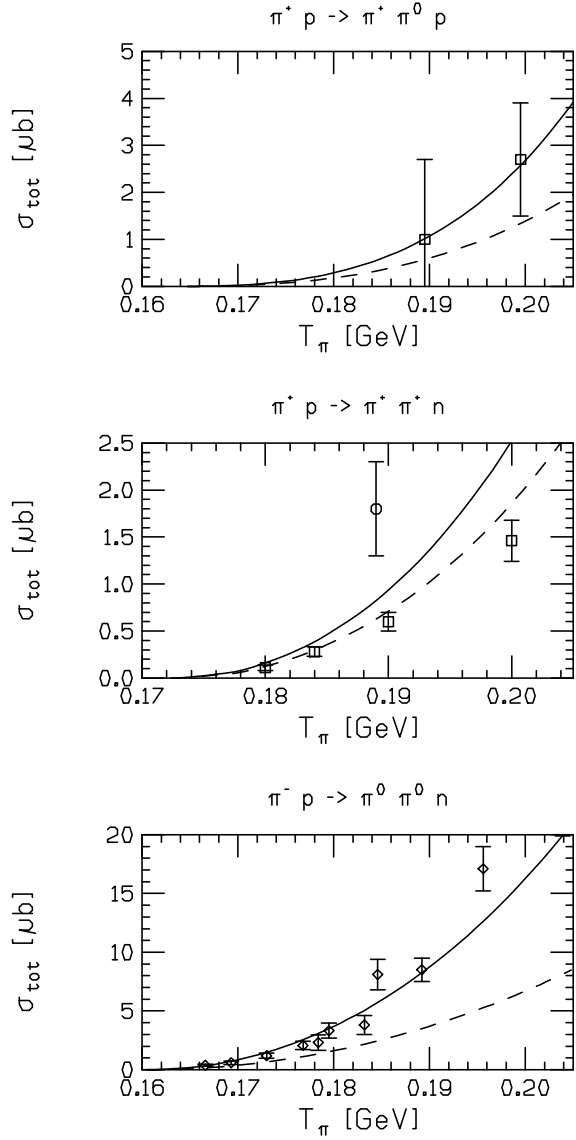}}
    \vspace{1cm}
   \centerline{\parbox{11cm}{\caption{\label{fig7}
Threshold cross section for $\pi^+ p \to \pi^+ \pi^0 p$,
$\pi^+ p \to \pi^+ \pi^+ n$  and  $\pi^- p \to \pi^0 \pi^0 n$.
  }}}
\end{figure}

\begin{figure}[t]
   \vspace{0.5cm}
   \epsfxsize=7.5cm
   \centerline{\epsffile{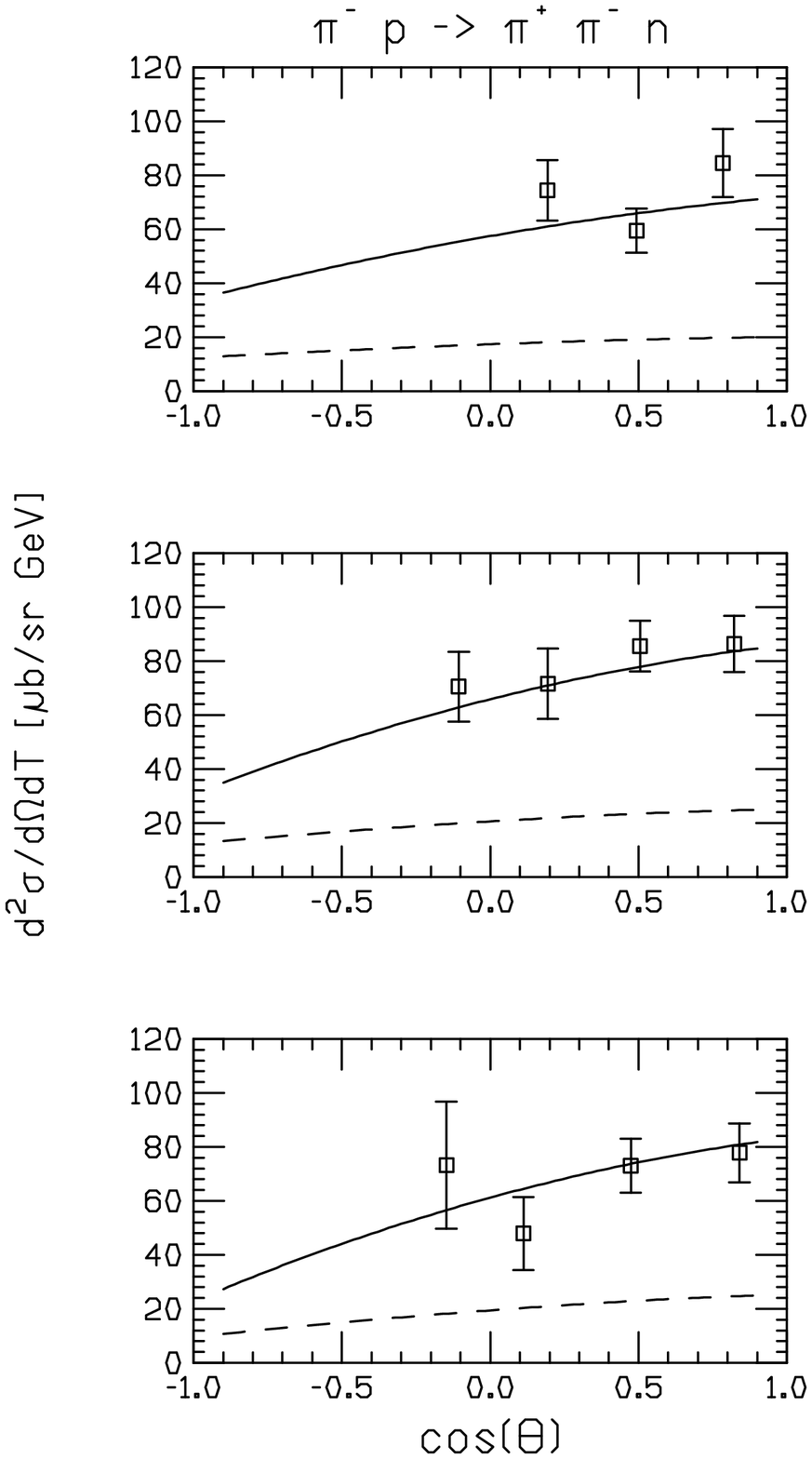}}
    \vspace{1cm}
   \centerline{\parbox{12cm}{\caption{\label{fdd1}
Double differential cross section at $\protect{\sqrt{s} = 1.242}$~GeV 
in comparison to the data of \protect{\cite{manl}} for $T_{\pi^+} = \omega_1-
M_\pi = 6.5,10.95, 15.35\,$MeV in the upper, middle and lower panel,
respectively. }}}
\end{figure}

\begin{figure}[t]
   \vspace{0.5cm}
   \epsfxsize=7.5cm
   \centerline{\epsffile{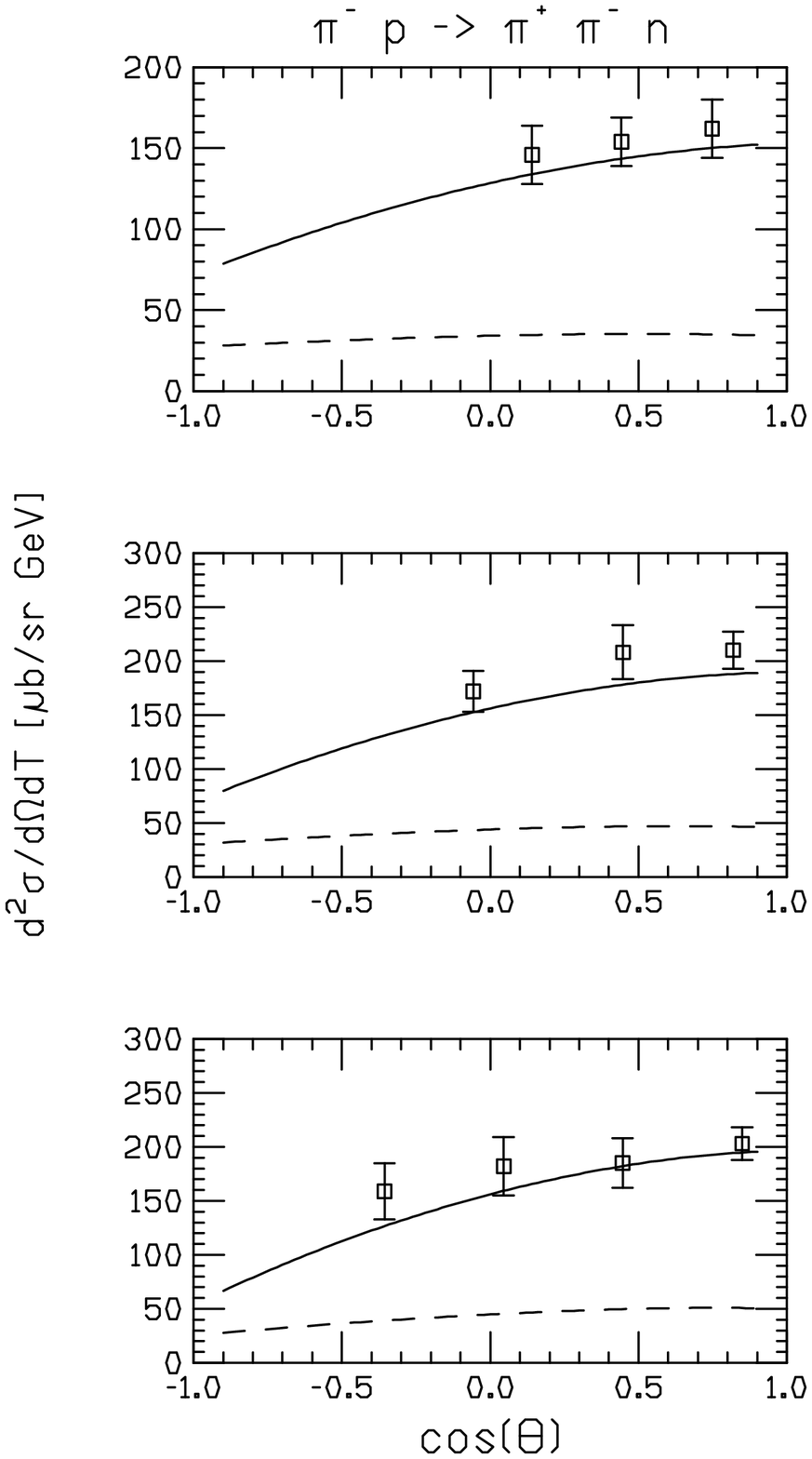}}
    \vspace{1cm}
   \centerline{\parbox{12cm}{\caption{\label{fdd2}
Double differential cross section at $\protect{\sqrt{s} = 1.262}$~GeV 
in comparison to the data of \protect{\cite{manl}} for $T_{\pi^+} =
\omega_1-M_\pi = 10.4, 17.5, 24.6\,$MeV in the upper, middle and lower panel,
respectively. }}}
\end{figure}

\begin{figure}[t]
   \vspace{0.5cm}
   \epsfxsize=7.5cm
   \centerline{\epsffile{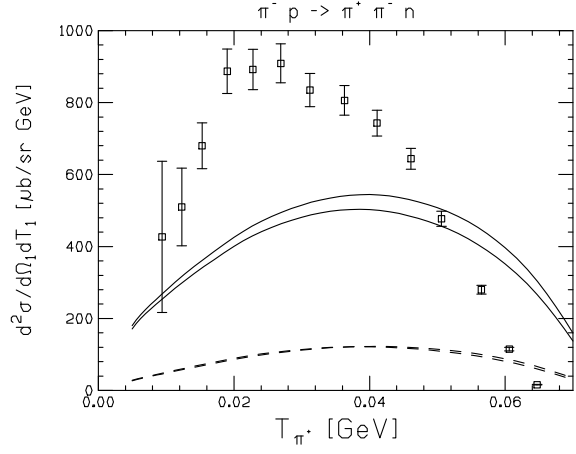}}
    \vspace{1cm}
   \centerline{\parbox{12cm}{\caption{\label{fdd3}
Double differential cross section for $\pi^- p \to \pi^+ \pi^- n$
in comparison to the data of \protect{\cite{mue}} at $\protect{\sqrt{s} =
1.301\,}$GeV with $0.03 < x_1 < 0.35$. $T_{\pi^+}= \omega_1 -M_\pi $ is the
kinetic energy of the emitted $\pi^+$ in the CMS.
  }}}
\end{figure}

\begin{figure}[t]
   \vspace{0.5cm}
   \epsfxsize=7.5cm
   \centerline{\epsffile{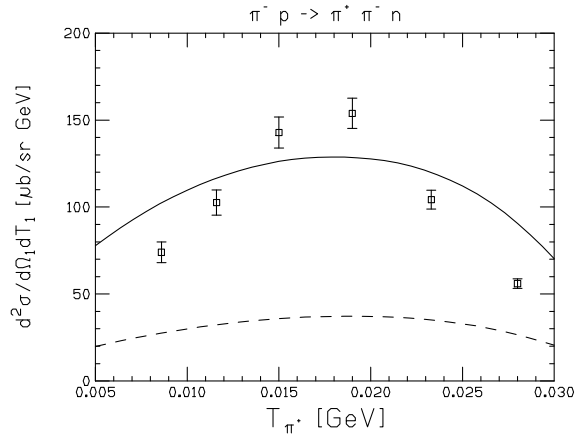}}
    \vspace{1cm}
   \centerline{\parbox{12cm}{\caption{\label{fdd4}
Double differential cross section for $\pi^- p \to \pi^+ \pi^- n$
in comparison to the data of \protect{\cite{boh}} at the incoming pion
energy $T_{\pi^-} = 218\,$MeV at $x_1 = 0.19$. $T_{\pi^+}= \omega_1 -M_\pi $ is
the kinetic energy of the emitted $\pi^+$ in the CMS.
  }}}
\end{figure}

\begin{figure}[t]
   \vspace{0.5cm}
   \epsfxsize=11cm
   \centerline{\epsffile{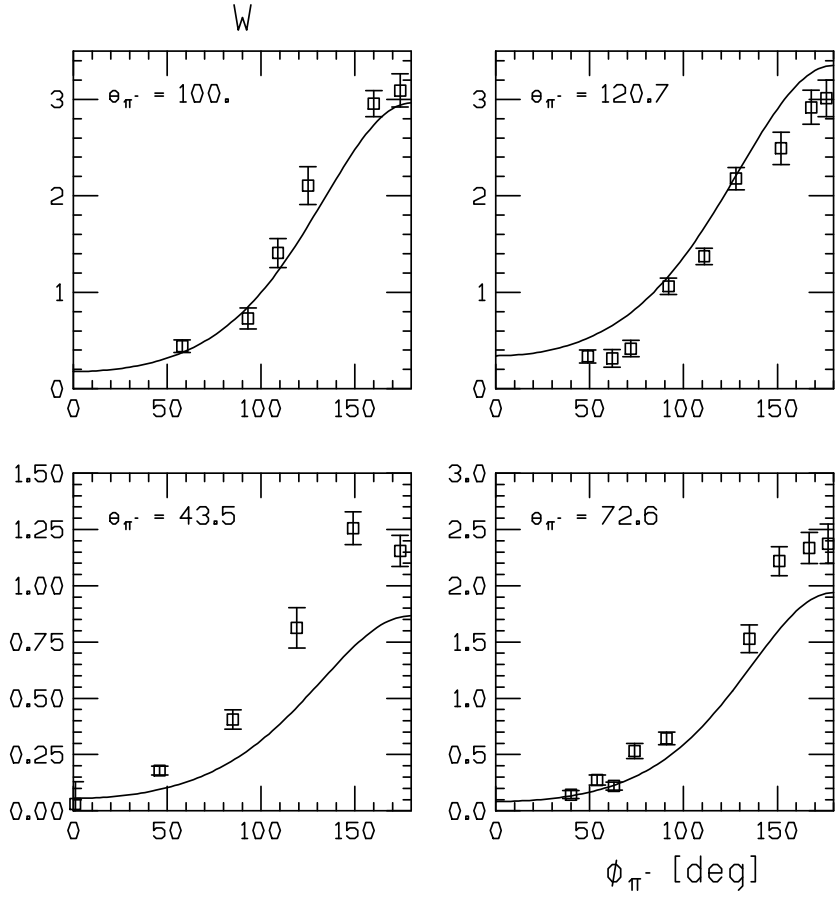}}
    \vspace{1cm}
   \centerline{\parbox{12cm}{\caption{\label{fw1}
Angular correlation function $W$ for $\pi^- p \to \pi^+ \pi^- n$
in comparison to the data of \protect{\cite{mue}} for the incoming pion
energy $T_{\pi^-} = 284\,$MeV at $|\vec q_1|  = 127\,$MeV and 
$\theta_2$ as given in the various panels.
  }}}
\end{figure}

\begin{figure}[t]
   \vspace{0.5cm}
   \epsfxsize=13cm
   \centerline{\epsffile{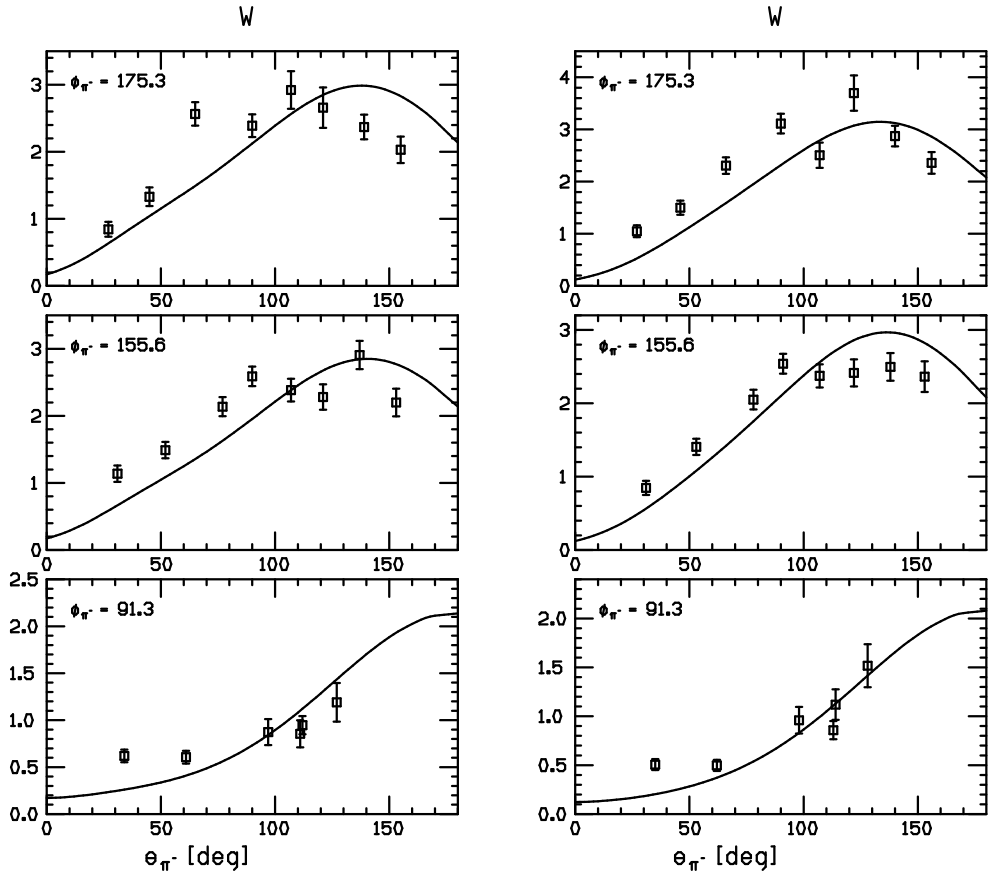}}
    \vspace{-1cm}
   \centerline{\parbox{12.5cm}{\caption{\label{fw2}
Angular correlation function $W$ for $\pi^- p \to \pi^+ \pi^- n$
in comparison to the data of \protect{\cite{mue}} for the incoming pion
energy $T_{\pi^-} = 284\,$MeV at $|\vec q_1| = 60.5\,$MeV and  $| \vec q_1| 
= 86\,$MeV (left and right panels, respectively) at fixed
$\phi_2$ as given in the various panels.
  }}}
\end{figure}

\end{document}